\documentclass[12pt, draftcls, onecolumn, journal]{IEEEtran}
\usepackage{amsmath,amssymb,latexsym}
\usepackage{bm}
\usepackage{graphicx}
\usepackage{subfigure}
\usepackage{cite}
\usepackage{algorithmic}
\usepackage{algorithm}
\usepackage{epsfig}

\title{Optimal Bandwidth and Power Allocation for \\ Sum Ergodic
Capacity under Fading Channels \\ in Cognitive Radio Networks}
\author{Xiaowen~Gong,~\IEEEmembership{Student Member,~IEEE},
Sergiy~A.~Vorobyov,~\IEEEmembership{Senior Member,~IEEE}, and
Chintha Tellambura, ~\IEEEmembership{Senior Member,~IEEE}
\thanks{This work was supported in parts by research grants from the
Natural Science and Engineering Research Council (NSERC) of Canada
and Alberta Ingenuity New Faculty Award.}
\thanks{The authors are with the Department of Electrical and Computer
Engineering, University of Alberta, 9107-116 St., Edmonton,
Alberta, T6G~2V4 Canada. The contacting emails are \{xgong2,
vorobyov, chintha\}@ece.ualberta.ca.}
\thanks{{\bf Corresponding author:} Sergiy A. Vorobyov, Dept. of \
Electrical and Computer Engineering, University of Alberta,
9107-116~St., Edmonton, Alberta, T6G 2V4, Canada; Phone: +1 (780)
492 9702, Fax: +1 (780) 492 1811.} }

\begin{document}

\maketitle

\begin{abstract}

This paper studies optimal bandwidth and power allocation in a
cognitive radio network where multiple secondary users (SUs) share
the licensed spectrum of a primary user (PU) under fading channels
using the frequency division multiple access scheme. The sum
ergodic capacity of all the SUs is taken as the performance metric
of the network. Besides all combinations of the peak/average
transmit power constraints at the SUs and the peak/average
interference power constraint imposed by the PU, total bandwidth
constraint of the licensed spectrum is also taken into account.
Optimal bandwidth allocation is derived in closed-form for any
given power allocation. The structures of optimal power
allocations are also derived under all possible combinations of
the aforementioned power constraints. These structures indicate
the possible numbers of users that transmit at nonzero power but
below their corresponding peak powers, and show that other users
do not transmit or transmit at their corresponding peak power.
Based on these structures, efficient algorithms are developed for
finding the optimal power allocations.

\end{abstract}


\section{Introduction}

Cognitive radio is a promising technology for improving spectrum
utilization in wireless communications systems \cite{Mitola}. A
secondary user (SU) in a cognitive radio network is allowed to
access the licensed spectrum allocated to a primary user (PU) if
the spectrum is not utilized by the PU. Such a spectrum sharing
strategy, which is referred to as \textit{spectrum overlay} or
\textit{opportunistic spectrum access} (OSA) \cite{Haykin},
requires correct detection of spectrum opportunities by the SU.
Existing works on spectrum overlay have mainly studied spectrum
sensing and access policies at the medium access control (MAC)
layer \cite{Zhao1}-\cite{Chang}. An alternative strategy, which is
known as \textit{spectrum underlay} \cite{Zhao2}--\cite{Zheng},
enables the PU and the SU to transmit simultaneously, provided
that the received interference power by the PU is below a
prescribed threshold level. A number of works have recently
studied information theoretic limits for resource allocation in
the context of spectrum underlay.

In \cite{Ghasemi}, the optimal power allocation which aims at
maximizing the ergodic capacity achieved by an SU is derived for
various channel fading models subject to the peak interference
power (PIP) constraint  or average interference power (AIP)
constraint imposed by a PU. In \cite{Musavian}, the authors derive
the optimal power allocation for the ergodic capacity, outage
capacity, and minimum-rate capacity of an SU under both the PIP
and AIP constraints from a PU. The ergodic capacity, delay-limited
capacity, and outage capacity of an SU is studied in \cite{Kang}
under different combinations of the peak transmit power (PTP)
constraint or average transmit power (ATP) constraint at the SU
and the PIP constraint or AIP constraint from a PU. However, all
the papers mentioned above consider the setup of a single SU. The
most recent work \cite{Zhang1} studies a cognitive radio network
of multiple SUs under multiple access channel and broadcast
channel models, where the optimal power allocation is derived to
achieve the maximum sum ergodic capacity of the SUs subject to
various mixed transmit and interference power constraints. The
optimality conditions for the dynamic time division multiple
access scheme are also derived.

In this paper, we focus on a cognitive radio network where
multiple SUs share the licensed spectrum of a PU using the
frequency division multiple access (FDMA) scheme. The sum ergodic
capacity of the SUs, which is a relevant network performance
metric for delay-tolerant traffics, is studied. Besides the
transmit power constraints at the SUs and the interference power
constraint imposed by the PU, which are also considered in
\cite{Ghasemi}-\cite{Zhang1}, we also take into account the total
bandwidth constraint of the shared spectrum. Such study is
motivated by the fact observed for a number of different
applications that joint bandwidth and power allocation can
significantly improve the performance of systems with limited both
individual (power) and public (bandwidth) resources
\cite{Yu}--\cite{Gong2}. Thus, in this paper, instead of
conventional fixed and equal bandwidth allocation used in FDMA, we
investigate dynamic and unequal bandwidth allocation, where the
bandwidth allocation varies for different SUs at different channel
fading states. Moreover, different from the existing works
\cite{Ghasemi}-\cite{Zhang1}, all combinations of the transmit
power constraints and the interference power constraints are
considered, including both PTP and ATP constraints combined with
both PIP and AIP constraints.

We first derive the optimal bandwidth allocation for any given
power allocation in any channel fading state, which results in
equivalent problems that only involve power allocation. Using the
convexity of the resultant power allocation problems, we apply
dual decomposition which transforms these problems into equivalent
dual problems, where each dual function involves a power
allocation subproblem associated with a specific channel fading
state. The dual problems can be solved using standard subgradient
algorithms. For the power allocation subproblem under all
combinations of the power constraints, we derive the structures of
the optimal power allocations. These structures indicate the
possible numbers of users that transmit at nonzero power but below
their corresponding peak powers, and show that other users do not
transmit or transmit at their corresponding peak power. Based on
these structures, we develop algorithms for finding the optimal
power allocations in each channel fading state.

The rest of the paper is organized as follows.
Section~\ref{sec_mdl} summarizes the system model and formulates
corresponding sum ergodic capacity maximization problems.
Section~\ref{sec_oba} derives the optimal bandwidth allocation for
the problems formulated in Section~\ref{sec_mdl} subject to the
bandwidth constraint. Section~\ref{sec_opa} obtains the optimal
power allocations from the resultant problems in
Section~\ref{sec_oba} under all combinations of the transmit power
constraints and interference power constraints. Numerical results
for the maximum sum ergodic capacity under different combinations
of the power constraints and the bandwidth constraint are shown in
Section~\ref{sec_sim}. Section~\ref{sec_con} concludes the paper.

\section{System Model}\label{sec_mdl}

Consider a cognitive radio network of $N$ SUs and one PU. The PU
occupies a spectrum of bandwidth $W$ for its transmission, while
the same spectrum is shared by the SUs. The spectrum is assumed to
be divided into distinct and nonoverlapping flat fading channels
with different bandwidth, so that the SUs share the spectrum
through FDMA to avoid interferences with each other. The channel
power gains between the $i$th SU transmitter (SU-Tx) and the $i$th
SU receiver (SU-Rx) and between the $i$th SU-Tx and the PU
receiver (PU-Rx) are denoted by $h_i$ and $g_i$, respectively. The
channel power gains, i.e., $\bm{g} \triangleq [g_1 \ g_2 \ \cdots
\ g_N]$ and $\bm{h} \triangleq [h_1 \ h_2 \ \cdots \ h_N]$, are
assumed to be drawn from an ergodic and stationary vector random
process. We further assume that full channel state information
(CSI), i.e., the joint probability density function (PDF) of the
channel power gains and the instantaneous channel power gains, are
known at the SUs.\footnote{Note that the full CSI assumption is
typical in the context of cognitive radio and is also made in
other works such as \cite{Ghasemi}-\cite{Zhang1}. Indeed, under
this assumption we aim at investigating the information-theoretic
limits on the sum ergodic capacity.} The noise at each SU-Rx plus
the interference from the PU transmitter (PU-Tx) is assumed to be
additive white Gaussian noise (AWGN) with unit power spectral
density (PSD).

We denote the transmit power of the $i$th SU-Tx and the channel
bandwidth allocated to the $i$th SU-Tx as $p_i(\bm{g},\bm{h})$ and
$w_i(\bm{h},\bm{g})$, respectively, for the instantaneous channel
power gains $\bm{g}$ and $\bm{h}$. Then the total bandwidth
constraint can be expressed as
\begin{equation}\label{bc}
\sum^{N}_{i=1} w_i(\bm{h},\bm{g}) \leq W, \ \forall \ \bm{h},\bm{g}.
\end{equation}
The PTP constraints are given by
\begin{equation}\label{pc_pk}
p_i(\bm{h},\bm{g}) \leq P^{pk}_i, \ \forall \ i, \bm{h}, \bm{g}
\end{equation}
where $P^{pk}_i$ denotes the maximum peak transmit power of the
$i$th SU-Tx. The PIP constraint is given by
\begin{equation}\label{ic_pk}
\sum^{N}_{i=1} g_ip_i(\bm{h},\bm{g}) \leq Q^{pk}, \ \forall \
\bm{h}, \bm{g}
\end{equation}
where $Q^{pk}$ denotes the maximum peak interference power allowed
at the PU-Rx. The ATP constraints are given by
\begin{equation}\label{pc_av}
\textrm{E}\left\{p_i(\bm{h},\bm{g})\right\} \leq P^{av}_i, \
\forall \ i
\end{equation}
where the expectation is taken over $\bm{h}$ and $\bm{g}$, and
$P^{av}_i$ denotes the maximum average transmit power of the $i$th
SU-Tx. The AIP constraint is given by
\begin{equation}\label{ic_av}
\textrm{E}\left\{\sum^{N}_{i=1} g_ip_i(\bm{h},\bm{g})\right\}
\leq Q^{av}
\end{equation}
where $Q^{av}$ denotes the maximum average interference power
allowed at the PU-Rx.

The objective is to maximize the sum ergodic capacity of the SUs,
which can be written as
\begin{equation}\label{sc}
\max_{\{w_i(\bm{h},\bm{g}), p_i(\bm{h},\bm{g})\} \in \mathcal{F}}
\textrm{E}\left\{\sum^{N}_{i=1} w_i(\bm{h}, \bm{g}) \log\left(1 +
\frac{h_ip_i(\bm{h},\bm{g})}{w_i(\bm{h},\bm{g})} \right) \right\}
\end{equation}
where $\mathcal{F}$ is a feasible set specified by the bandwidth
constraints \eqref{bc} and a particular combination of the
transmit power constraints $\{\eqref{pc_pk}, \eqref{pc_av}\}$ and
the interference power constraints $\{\eqref{ic_pk},
\eqref{ic_av}\}$. Note that the constraints on nonnegativity of
the bandwidth and power allocations, i.e., $w_i(\bm{h},\bm{g})
\geq 0$ and $p_i(\bm{h},\bm{g}) \geq 0$, $\forall i, \bm{h},
\bm{g}$, are natural and, thus, omitted through out the paper for
brevity.

It can be shown that the objective function of the problem
\eqref{sc} is concave with respect to $\{w_i(\bm{h},\bm{g}),
p_i(\bm{h},\bm{g})\}$, $\forall i, \bm{h}, \bm{g}$. It can also be
seen that the bandwidth and power constraints
\eqref{bc}--\eqref{ic_av} are linear and, thus, convex. Therefore,
the sum ergodic capacity maximization problem \eqref{sc} under
different combinations of the constraints
\eqref{bc}--\eqref{ic_av} is a convex optimization problem.

\section{Optimal Bandwidth Allocation}\label{sec_oba}

Given that the power allocation $p_i(\bm{h},\bm{g})$, $\forall i,
\bm{h}, \bm{g}$, is fixed, the maximum sum ergodic capacity can be
expressed as $\textrm{E}\{f_0(\bm{h},\bm{g})\}$, where
$f_0(\bm{h},\bm{g})$ is given by
\begin{subequations}
\begin{eqnarray}
f_0(\bm{h},\bm{g}) \triangleq & &  \max_{\{w_i(\bm{h},\bm{g})\}}
\sum^{N}_{i=1} G_i\left(w_i(\bm{h},\bm{g})\right)\label{ba1}\\
& & \textrm{s.t.} \ \sum^{N}_{i=1} w_i(\bm{h},\bm{g}) \leq W
\label{ba2} \\
\end{eqnarray}
\end{subequations}
where $G_i(w_i(\bm{h},\bm{g})) \triangleq w_i(\bm{h},\bm{g}) \log
\left(1 + h_ip_i(\bm{h},\bm{g}) / w_i(\bm{h},\bm{g}) \right)$ is
an increasing and concave function of $w_i(\bm{h},\bm{g})$. The
problem \eqref{ba1}--\eqref{ba2} is similar to the classical
water-filling power allocation problem. Thus, the optimal solution
of the problem \eqref{ba1}--\eqref{ba2}, denoted by
$\{w^{\prime}_i(\bm{h},\bm{g})\}$, must satisfy
\begin{equation}\label{ba3}
\frac{\partial G_i(w_i(\bm{h}, \bm{g}))}{\partial w_i(\bm{h},
\bm{g})}\bigg|_{w_i(\bm{h}, \bm{g}) = w^{\prime}_i(\bm{h},
\bm{g})} = \frac{\partial G_j(w_j(\bm{h}, \bm{g}))}{\partial
w_j(\bm{h}, \bm{g})} \bigg|_{w_j(\bm{h}, \bm{g}) =
w^{\prime}_j(\bm{h},\bm{g})}, \ \forall \ i \neq j.
\end{equation}
Since we have
\begin{equation}\label{ba4}
\begin{split}
\frac{\partial G_i(w_i(\bm{h},\bm{g}))}{\partial w_i(\bm{h},
\bm{g})} \bigg|_{w_i(\bm{h}, \bm{g}) \!\!=\! w^{\prime}_i(\bm{h},
\bm{g})} &= \log \left(1 \!+\! \frac{h_ip_i(\bm{h},
\bm{g})}{w^{\prime}_i(\bm{h}, \bm{g})}\right) \!-\!
\frac{h_ip_i(\bm{h},\bm{g})}{w^{\prime}_i(\bm{h},\bm{g}) \!+\!
h_ip_i(\bm{h},\bm{g})} \!=\! Y \left(\frac{h_ip_i(\bm{h},
\bm{g})}{w^{\prime}_i(\bm{h}, \bm{g})}\right)
\end{split}
\end{equation}
where $Y(x) \triangleq \log(1 + x) - x/(1 + x)$ is a monotonically
increasing function, we can obtain from \eqref{ba3} that
\begin{equation}\label{ba5}
\frac{h_ip_i(\bm{h}, \bm{g})}{w^{\prime}_i(\bm{h}, \bm{g})} =
\frac{h_jp_j(\bm{h}, \bm{g})}{w^{\prime}_j(\bm{h}, \bm{g})}, \quad
\forall \ i \neq j
\end{equation}
It follows from \eqref{ba2} that at optimality we have
$\sum^{N}_{i=1} w^{\prime}_i(\bm{h},\bm{g}) = W$. Furthermore,
using \eqref{ba5}, we can obtain that
\begin{equation}\label{ba6}
w^{\prime}_i(\bm{h}, \bm{g}) = W \frac{h_ip_i(\bm{h},
\bm{g})}{\sum^{N}_{i=1} h_ip_i(\bm{h}, \bm{g})}.
\end{equation}
Substituting the optimal $w_i(\bm{h},\bm{g})$ given by \eqref{ba6}
into \eqref{sc}, we can equivalently rewrite \eqref{sc} as
\begin{equation}\label{sc_eq}
\max_{\{p_i(\bm{h},\bm{g})\} \in \mathcal{F^{\prime}}}
\textrm{E}\left\{W \log \left(1 + \frac{\sum^{N}_{i=1}
h_ip_i(\bm{h},\bm{g})}{W}\right)\right\}
\end{equation}
where $\mathcal{F^{\prime}}$ is a feasible set specified only by a
particular combination of the power constraints $\{\eqref{pc_pk},
\eqref{ic_pk}, \eqref{pc_av}, \eqref{ic_av}\}$. Therefore, the
optimal power allocation obtained from the problem \eqref{sc} and
denoted by $\{p^{\ast}_i(\bm{h},\bm{g})\}$, can also be obtained
by solving the equivalent problem \eqref{sc_eq}. Then the optimal
bandwidth allocation obtained from the problem \eqref{sc} and
denoted by $\{w^{\ast}_i(\bm{h},\bm{g})\}$, can be found as
\begin{equation}\label{ba_sln}
w^{\ast}_i(\bm{h}, \bm{g}) = W \frac{h_ip^{\ast}_i(\bm{h},
\bm{g})}{\sum^{N}_{i=1} h_ip^{\ast}_i(\bm{h}, \bm{g})}.
\end{equation}

\section{Optimal Power Allocation}\label{sec_opa}

In this section, we study the optimal power allocation obtained
from the problem \eqref{sc_eq} with $\mathcal{F^{\prime}}$
specified by different combinations of the power constraints.

\subsection{Peak transmit power with peak interference
power constraints}\label{pa_pp}

Consider $\mathcal{F^{\prime}} = \{\textrm{the \ constraints
\eqref{pc_pk} and \eqref{ic_pk}}\}$. Then the optimal value of the
problem \eqref{sc_eq} can be expressed as
$\textrm{E}\left\{f_1(\bm{h},\bm{g})\right\}$, where
$f_1(\bm{h},\bm{g})$ is given by
\begin{subequations}
\begin{eqnarray}
f_1(\bm{h}, \bm{g}) \triangleq & & \max_{\{p_i(\bm{h}, \bm{g})\}}
W \log\left(1+\frac{\sum^{N}_{i=1} h_ip_i(\bm{h},\bm{g})}
{W} \right) \label{pa_pp1} \\
& & \textrm{s.t.} \ p_i(\bm{h},\bm{g}) \leq P^{pk}_i, \
\forall \ i \label{pa_pp2} \\
& & \quad \sum^{N}_{i=1} g_i p_i (\bm{h}, \bm{g}) \leq Q^{pk}.
\label{pa_pp3}
\end{eqnarray}
\end{subequations}
For brevity, we drop the dependence on $\bm{h}$ and $\bm{g}$ that
specifies instantaneous channel power gains. Also let
$\{p^{\ast}_i\}$ denote the optimal solution of the problem
\eqref{pa_pp1}--\eqref{pa_pp3}. Introducing $q_i \triangleq
g_ip_i$, the problem \eqref{pa_pp1}--\eqref{pa_pp3} can be
equivalently rewritten as
\begin{subequations}
\begin{eqnarray}
& & \max_{\{q_i\}} \sum^{N}_{i=1} \frac{h_i}{g_i}q_i
\label{pa_pp4} \\
& & \textrm{s.t.} \ q_i \leq g_iP^{pk}_i, \ \forall \ i
\label{pa_pp5} \\
& & \quad \sum^{N}_{i=1} q_i \leq Q^{pk}.\label{pa_pp6}
\end{eqnarray}
\end{subequations}
Let $\{q^{\ast}_i\}$ denote the optimal solution of the problem
\eqref{pa_pp4}--\eqref{pa_pp6} and $(s_1, s_2, \cdots, s_N)$
denote a permutation of the SU indexes such that $h_{s_1}/g_{s_1}
> h_{s_2}/g_{s_2} > \cdots > h_{s_N}/g_{s_N}$. It is assumed that
$h_i/g_i \neq h_j/g_j$, $\forall i \neq j$, since $h_i$, $g_i$,
$h_j$, and $g_j$ are drawn from a continuous-valued random
process. Then the following lemma is in order.

\textbf{Lemma~1:} {\it There exists $k$, $1 \leq k \leq N$, such
that $q^{\ast}_{s_i} = g_{s_i}P^{pk}_{s_i}$, $\forall i$, $1 \leq
i \leq k-1$, $0 < q^{\ast}_{s_k} \leq g_{s_k}P^{pk}_{s_k}$, and
$q^{\ast}_{s_i} = 0$, $\forall i$, $k+1 \leq i \leq N$.}

\textit{Proof:} \ Let $q^{\ast}_{s_j} > 0$ for some $j$ and let $l
< j$ for some $l$. First we prove that $q^{\ast}_{s_l} =
g_{s_l}P^{pk}_{s_l}$ by contradiction. If $q^{\ast}_{s_l} <
g_{s_l}P^{pk}_{s_l}$, then we can always find $\Delta q > 0$ and
define a feasible solution $\{q^{\prime}_{s_i}\}$ of the problem
\eqref{pa_pp4}--\eqref{pa_pp6} $q^{\prime}_{s_j} \triangleq
q^{\ast}_{s_j} - \Delta q, q^{\prime}_{s_l} \triangleq
q^{\ast}_{s_l} + \Delta q, q^{\prime}_{s_i} \triangleq
q^{\ast}_{s_i}, \forall i, i \neq j, i \neq l$ such that the
objective function in \eqref{pa_pp4} achieves larger value for
$\{q^{\prime}_{s_i}\}$ than for the optimal solution
$\{q^{\ast}_i\}$, since we have
\begin{equation}\label{pa_pp7}
\sum^{N}_{i=1} \frac{h_{s_i}}{g_{s_i}}q^{\prime}_{s_i} -
\sum^{N}_{i=1} \frac{h_{s_i}}{g_{s_i}} q^{\ast}_{s_i} = \left(
\frac{h_{s_l}}{g_{s_l}} - \frac{h_{s_j}}{g_{s_j}} \right) \Delta q
> 0.
\end{equation}
Therefore, it contradicts the fact that $\{q^{\ast}_{s_i}\}$ is
the optimal solution of the problem
\eqref{pa_pp4}--\eqref{pa_pp6}.

Let $q^{\ast}_{s_j} < g_{s_j}P^{pk}_{s_j}$ for some $j$ and let $l
> j$ for some $l$. Using the result obtained above, it can be
proved also by contradiction that $q^{\ast}_{s_l} = 0$. This
completes the proof. $\hfill\square$

Lemma~1 shows that for the optimal power allocation under the
constraints \eqref{pc_pk} and \eqref{ic_pk}, there exists at most
one user that transmits at nonzero power and below its peak power,
while any other user either does not transmit or transmits at its
peak power.

Note that either the constraints \eqref{pa_pp5} or the constraint
\eqref{pa_pp6} must be active at optimality. Using the structure
of $\{q^{\ast}_i\}$ given in Lemma~1, $k$ can be found by
Algorithm~1.
\begin{algorithm}
\caption{Algorithm for finding $k$ in Lemma~1} \label{pa_pp_al}
\begin{algorithmic}
\STATE \textbf{Initialize:} $k = 1$

\WHILE{$\sum^{k}_{i=1} g_{s_i}P^{pk}_{s_i} < Q^{pk}$ and $k \leq
N-1$}

\STATE $k = k + 1$

\ENDWHILE

\STATE \textbf{Output:} $k$
\end{algorithmic}
\end{algorithm}

Since $p^{\ast}_{s_i} = q^{\ast}_{s_i} / g_{s_i}$, we obtain
\begin{equation}\label{pa_pp_sln}
p^{\ast}_{s_i} = \left\{ {\begin{array}{ll}
   P^{pk}_{s_i}, & 1 \leq i \leq k-1 \\
   \min\{P^{pk}_{s_i},(Q^{pk} - \sum^{k-1}_{i=1} g_{s_i}
   P^{pk}_{s_i}) / g_{s_i}\}, & i = k \\
   0, & k+1 \leq i \leq N. \\
\end{array}} \right.
\end{equation}
Note that for brevity, we say in this paper that
$\sum^{n}_{i=1}x_i = 0$ if $n = 0$ with a little abuse of
notation.

\subsection{Average transmit power with average interference
power constraints}\label{pa_aa}

Consider $\mathcal{F^{\prime}} = \{\textrm{the \ constraints
\eqref{pc_av} and \eqref{ic_av}}\}$. Then the dual function of the
problem \eqref{sc_eq} can be written as
\begin{equation}\label{pa_aa_dual}
f_2(\{\lambda_i\}, \mu) \triangleq \textrm{E} \left\{
f^{\prime}_2(\bm{h}, \bm{g})\right\} + \sum^{N}_{i=1} \lambda_i
P^{av}_i + \mu Q^{av}
\end{equation}
where $\{\lambda_i|1\leq i \leq N\}$ and $\mu$ are the nonnegative
dual variables associated with the corresponding constraints in
\eqref{pc_av} and \eqref{ic_av} and $f^{\prime}_2(\bm{h},\bm{g})$
is given by
\begin{equation}\label{pa_aa1}
f^{\prime}_2 (\bm{h}, \bm{g}) \triangleq \max_{\{p_i(\bm{h},
\bm{g})\}} W \log\left( 1 + \frac{\sum^{N}_{i=1} h_i p_i (\bm{h},
\bm{g})} {W} \right) - \sum^{N}_{i=1} \gamma_i p_i(\bm{h}, \bm{g})
\end{equation}
with $\gamma_i \triangleq \lambda_i + \mu g_i$. Let
$\{p^{\ast}_i\}$ denote the optimal solution of the problem
\eqref{pa_aa1}, where we drop the dependence on $\bm{h}$ and
$\bm{g}$ for brevity. Also let $F(\{p_i\})$ denote the objective
function in \eqref{pa_aa1}. If $p^{\ast}_i > 0$ for some $i$, the
following must hold
\begin{equation}\label{pa_aa2}
\frac{\partial F(\{p_i\})}{\partial p_i}\bigg|_{\{p_i\} =
\{p^{\ast}_i\}} = \frac{h_i}{1 + \sum^{N}_{i=1} h_i p^{\ast}_i /
W} - \gamma_i = 0.
\end{equation}
Then the following lemma is of interest.

\textbf{Lemma~2:} {\it If $h_i \leq \gamma_i$ for some $i$, then
$p^{\ast}_i = 0$.}

\textit{Proof:} \ If $p^{\ast}_j = 0$, $\forall j$, then
$p^{\ast}_i = 0$. If $p^{\ast}_j \neq 0$ for some $j$, it can be
seen that \eqref{pa_aa2} can not be satisfied since $h_i \leq
\gamma_i$. Thus, $p^{\ast}_i = 0$. $\hfill\square$

If $p^{\ast}_i = 0$ for some $i$, the following must hold
\begin{equation}\label{pa_aa3}
\frac{\partial F(\{p_i\})}{\partial p_i}\bigg|_{\{p_i\} =
\{p^{\ast}_i\}} = \frac{h_i}{1 + \sum^{N}_{i=1} h_i p^{\ast}_i /
W} - \gamma_i \leq 0.
\end{equation}
Then the next lemma is in order.

\textbf{Lemma~3:} {\it $p^{\ast}_i = 0$, $\forall i$, if and only
if $h_i \leq \gamma_i$, $\forall i$.}

\textit{Proof:} \ It can be seen from Lemma~2 that if $h_i \leq
\gamma_i$, $\forall i$, then $p^{\ast}_i = 0$, $\forall i$.
Moreover, it can be seen from \eqref{pa_aa3} that if $p^{\ast}_i =
0$, $\forall i$, then $h_i \leq \gamma_i$, $\forall i$.
$\hfill\square$

Let $(s_1, s_2, \cdots, s_N)$ denote a permutation of the SU
indexes such that $h_{s_1}/\gamma_{s_1} > h_{s_2}/\gamma_{s_2}
> \cdots > h_{s_N}/\gamma_{s_N}$. Then we can also prove the
following lemma.

\textbf{Lemma~4:} {\it There exists at most one $k$ such that
$p^{\ast}_k > 0$. Moreover, $k = s_1$.}

\textit{Proof:} \ We prove it by contradiction. It can be seen
from \eqref{pa_aa2} that if $p^{\ast}_i > 0$ and $p^{\ast}_j > 0$
for some $i \neq j$, the following must hold
\begin{equation}\label{pa_aa4}
\frac{h_i}{\gamma_i} = \frac{h_j}{\gamma_j}.
\end{equation}
Since $h_i$, $\gamma_i$, $h_j$, and $\gamma_j$ are independent
constants given in the problem \eqref{pa_aa1}, \eqref{pa_aa4} can
not be satisfied. Let $p^{\ast}_k > 0$ and $p^{\ast}_i = 0$,
$\forall i$, $i \neq k$. Then it follows from \eqref{pa_aa2} and
\eqref{pa_aa3} that the following must hold
\begin{equation}\label{pa_aa5}
\frac{h_k}{\gamma_k} \geq \frac{h_i}{\gamma_i}, \ \forall \ i \neq k.
\end{equation}
Therefore, we must have $k = s_1$. $\hfill\square$

Lemma~4 shows that for the optimal power allocation under the
constraints \eqref{pc_av} and \eqref{ic_av}, there exists at most
one user that transmits at nonzero power, while any other user
does not transmit.

\textit{Case 1:} \ Consider the case when $h_i \leq \gamma_i$,
$\forall i$. It follows from Lemma~3 that $p^{\ast}_i = 0$,
$\forall i$.

\textit{Case 2:} \ Consider the case when $h_i \leq \gamma_i$ does
not hold for some $i$. Using Lemma~4, let $p^{\ast}_k > 0$ and
$p^{\ast}_i = 0$, $\forall i$, $i \neq k$. Substituting
$\{p^{\ast}_i\}$ into \eqref{pa_aa2}, we have $p^{\ast}_{s_1} =
W(1 / \gamma_{s_1} -  1 / h_{s_1})$. Therefore, we obtain
\begin{equation}\label{pa_aa_sln}
p^{\ast}_{s_i} = \left\{ {\begin{array}{ll}
   W\left(1 / \left(\lambda_{s_1} + \mu g_{s_1}\right) - 1 /
   h_{s_1}\right), & i = 1 \\
   0, & 2 \leq i \leq N. \\
\end{array}} \right.
\end{equation}

\subsection{Peak transmit power with average interference
power constraints}\label{pa_pa}

Consider $\mathcal{F^{\prime}} = \{\textrm{the \ constraints
\eqref{pc_pk} and \eqref{ic_av}}\}$. Then the dual function of the
problem \eqref{sc_eq} can be written as
\begin{equation}\label{pa_pa_dual}
f_3(\mu) \triangleq \textrm{E}\left\{f^{\prime}_3(\bm{h},
\bm{g})\right\} + \mu Q^{av}
\end{equation}
where $\mu$ is the nonnegative dual variable associated with the
constraint \eqref{ic_av}, and $f^{\prime}_3(\bm{h},\bm{g})$ is
given by
\begin{subequations}
\begin{eqnarray}
f^{\prime}_3(\bm{h},\bm{g}) \triangleq & & \max_{\{p_i(\bm{h},
\bm{g})\}} W \log \left( 1 + \frac{\sum^{N}_{i=1} h_i p_i(\bm{h},
\bm{g})} {W} \right) - \mu \sum^{N}_{i=1} g_i p_i(\bm{h}, \bm{g})
\label{pa_pa1}\\
& & \textrm{s.t.} \ p_i(\bm{h}, \bm{g}) \leq P^{pk}_i, \quad
\forall \ i.\label{pa_pa2}
\end{eqnarray}
\end{subequations}
Let $\{p^{\ast}_i\}$ denote the optimal solution of the problem
\eqref{pa_pa1}--\eqref{pa_pa2} after dropping the dependence on
$\bm{h}$ and $\bm{g}$ for brevity. The following cases are of
interest.

\textit{Case 1:} \ Consider the case when $h_i \leq \mu g_i$,
$\forall i$. Since the problem \eqref{pa_pa1}--\eqref{pa_pa2}
without the constraints \eqref{pa_pa2} has the same form as the
problem \eqref{pa_aa1}, and $p_i = 0$, $\forall i$, satisfies the
constraint \eqref{pa_pa2}, it can be seen from Lemma~3 that
$p^{\ast}_i = 0$, $\forall i$.

\textit{Case 2:} \ Consider the case when $h_i \leq \mu g_i$ does
not hold for some $i$. The problem \eqref{pa_pa1}--\eqref{pa_pa2}
is equivalent to
\begin{subequations}
\begin{equation}\label{pa_pa3}
\max_{\{q_i\}} W \log\left(1+\frac{\sum^{N}_{i=1} h_iq_i / \mu
g_i}{W}\right) - \sum^{N}_{i=1} q_i
\end{equation}
\begin{equation}\label{pa_pa4}
\textrm{s.t.} \ q_i \leq \mu g_iP^{pk}_i, \ \forall \ i
\end{equation}
\end{subequations}
where $q_i \triangleq \mu g_ip_i$. Let $\{q^{\ast}_i\}$ denote the
optimal solution of the problem \eqref{pa_pa3}--\eqref{pa_pa4} and
$\quad$ $(s_1, s_2, \cdots, s_N)$ denote a permutation of the SU
indexes such that $h_{s_1}/\mu g_{s_1} > h_{s_2}/\mu g_{s_2} >
\cdots > h_{s_N}/\mu g_{s_N}$. Then the following lemma is in
order.

\textbf{Lemma~5:} {\it There exists $k$, $1 \leq k \leq N$, such
that $q^{\ast}_{s_i} = g_{s_i}P^{pk}_{s_i}$, $\forall i$, $1 \leq
i \leq k-1$, $0 < q^{\ast}_{s_k} \leq g_{s_k}P^{pk}_{s_k}$, and
$q^{\ast}_{s_i} = 0$, $\forall i$, $k+1 \leq i \leq N$.}

\textit{Proof:} \ Consider the following intermediate problem
\begin{subequations}
\begin{eqnarray}
& & \max_{\{q_i\}} \sum^{N}_{i=1} \frac{h_i}{\mu g_i} q_i
\label{pa_pa5} \\
& & \textrm{s.t.} \ q_i \leq \mu g_iP^{pk}_i, \ \forall \
i \label{pa_pa6} \\
& & \quad \sum^{N}_{i=1} q_i = Q \label{pa_pa7}
\end{eqnarray}
\end{subequations}
where $Q \triangleq \sum^{N}_{i = 1} q^{\ast}_i$ and it is unknown
since $\{q^{\ast}_i\}$ is unknown. Let $\{q^{\prime}_i\}$ denote
the optimal solution of the problem
\eqref{pa_pa5}--\eqref{pa_pa7}. If $\{q^{\prime}_i\} \neq
\{q^{\ast}_i\}$, we have $\sum^{N}_{i = 1} h_i q^{\prime}_i / \mu
g_i \geq \sum^{N}_{i = 1} h_i q^{\ast}_i / \mu g_i$ since
$\{q^{\ast}_i\}$ is a feasible solution of the problem
\eqref{pa_pa5}--\eqref{pa_pa7}. Then we have
\begin{equation}\label{pa_pa8}
F \left( \{q^{\prime}_i\} \right) - F \left( \{q^{\ast}_i\}
\right) = W \log \left( 1 + \frac{\sum^{N}_{i=1} h_i q^{\prime}_i
/ \mu g_i}{W} \right) - W \log \left( 1 + \frac{\sum^{N}_{i=1} h_i
q^{\ast}_i / \mu g_i}{W}\right) \geq 0
\end{equation}
where $F(\{q_i\})$ denotes the objective function in the problem
\eqref{pa_pa3}--\eqref{pa_pa4}. Since $\{q^{\prime}_i\}$ is a
feasible solution of the problem \eqref{pa_pa3}--\eqref{pa_pa4},
it contradicts the fact that $\{q^{\ast}_i\}$ is the optimal
solution of the problem \eqref{pa_pa3}--\eqref{pa_pa4}. Therefore,
it must be true that $\{q^{\prime}_i\} = \{q^{\ast}_i\}$.

It can be seen from the constraints \eqref{pa_pa4} that
$\sum^{N}_{i = 1} q^{\prime}_i = \sum^{N}_{i = 1} q^{\ast}_i = Q
\leq \sum^{N}_{i = 1} \mu g_iP^{pk}_i$. Then the problem
\eqref{pa_pa5}--\eqref{pa_pa7} is equivalent to the following
problem
\begin{subequations}
\begin{eqnarray}
& & \max_{\{q_i\}} \sum^{N}_{i=1} \frac{h_i}{\mu g_i}q_i
\label{pa_pa9} \\
& & \textrm{s.t.} \ q_i \leq \mu g_iP^{pk}_i, \ \forall \ i
\label{pa_pa10} \\
& & \quad \sum^{N}_{i=1} q_i \leq Q \label{pa_pa11}
\end{eqnarray}
\end{subequations}
since the constraint \eqref{pa_pa11} is active at optimality.
Therefore, the problem \eqref{pa_pa3}--\eqref{pa_pa4} is
equivalent to the problem \eqref{pa_pa9}--\eqref{pa_pa11}. Since
the problem \eqref{pa_pa9}--\eqref{pa_pa11} is similar to the
problem \eqref{pa_pp4}--\eqref{pa_pp6} in Section \ref{pa_pp}, we
conclude that $\{q^{\ast}_i\}$ has the same structure as that
given in Lemma~1. $\hfill\square$

The result of Lemma~5 is similar to that of Lemma~1. Specifically,
it shows that for the optimal power allocation under the
constraints \eqref{pc_pk} and \eqref{ic_av}, there exists at most
one user that transmits at nonzero power and below its peak power,
while any other user either does not transmit or transmits at its
peak power.

Using Lemma~5, let $q^{\ast}_{s_i} = \mu g_{s_i}P^{pk}_{s_i}$,
$\forall i$, $1 \leq i \leq k-1$, $0 < q^{\ast}_{s_k} \leq \mu
g_{s_i}P^{pk}_{s_i}$, and $q^{\ast}_{s_i} = 0$, $\forall i$, $k+1
\leq i \leq N$. Then we only need to find $k$ and $q^{\ast}_{s_k}$
to determine $\{q^{\ast}_i\}$.

Consider the case when $0 < q^{\ast}_{s_k} < \mu g_{s_k}
P^{pk}_{s_k}$, $1 \leq k \leq N$. Then the following must be true
\begin{equation}\label{pa_pa12}
\frac{\partial H(q_{s_k})}{\partial q_{s_k}}\bigg|_{q_{s_k} =
q^{\ast}_{s_k}} = \frac{h_{s_k} / \mu g_{s_k}}{1 + \left(
\sum^{N}_{i=1, i \neq k}h_{s_i}q^{\ast}_{s_i} / \mu g_{s_i} +
h_{s_k}q^{\ast}_{s_k} / \mu g_{s_k} \right) / W} - 1 = 0
\end{equation}
where
\begin{equation}\label{pa_pa13}
H(q_{s_k})  \triangleq W \log \left(1 + \frac{\sum^{N}_{i=1, i
\neq k}h_{s_i}q^{\ast}_{s_i} / \mu g_{s_i} + h_{s_k}q_{s_k} / \mu
g_{s_k} }{W} \right) - \sum^{N}_{i=1, i \neq k} q^{\ast}_{s_i} -
q_{s_k} .
\end{equation}
Substituting $\{q^{\ast}_{s_i}\}$ into \eqref{pa_pa12}, we obtain
$q^{\ast}_{s_k} = W(1 - \mu g_{s_k} / h_{s_k}) - \mu g_{s_k}
\sum^{k-1}_{i=1} h_{s_i} P^{pk}_{s_i} / h_{s_k}$. Since
$q^{\ast}_{s_k}$ must satisfy $0 < q^{\ast}_{s_k} < \mu
g_{s_i}P^{pk}_{s_i}$, it must be true that
\begin{equation}\label{pa_pa14}
\sum^{k - 1}_{i=1}h_{s_i}P^{pk}_{s_i} < W\left(\frac{h_{s_k}}{ \mu
g_{s_k}} - 1\right) < \sum^{k}_{i=1}h_{s_i}P^{pk}_{s_i}.
\end{equation}

Consider the case when $q^{\ast}_{s_k} = \mu g_{s_k}P^{pk}_{s_k}$,
$1 \leq k \leq N - 1$. Then the following must hold
\begin{equation}\label{pa_pa15}
\frac{\partial H(q_{s_k})}{\partial q_{s_k}}\bigg|_{q_{s_k} =
q^{\ast}_{s_k}} = \frac{h_{s_k} / \mu g_{s_k}}{1 + \left(
\sum^{N}_{i=1, i \neq k}h_{s_i}q^{\ast}_{s_i} / \mu g_{s_i} +
h_{s_k}q^{\ast}_{s_k} / \mu g_{s_k} \right) / W} - 1 \geq 0
\end{equation}
and
\begin{equation}\label{pa_pa16}
\frac{\partial H(q_{s_{k + 1}})}{\partial q_{s_{k +
1}}}\bigg|_{q_{s_{k + 1}} = q^{\ast}_{s_{k + 1}}} = \frac{h_{s_{k
+ 1}} / \mu g_{s_{k + 1}}}{1 + \left( \sum^{N}_{i=1, i \neq
k+1}h_{s_i}q^{\ast}_{s_i} / \mu g_{s_i} + h_{s_{k +
1}}q^{\ast}_{s_{k + 1}} / \mu g_{s_{k + 1}} \right) / W} - 1 \leq
0.
\end{equation}
Substituting $\{q^{\ast}_i\}$ into \eqref{pa_pa15} and
\eqref{pa_pa16}, we obtain
\begin{equation}\label{pa_pa17}
W\left(\frac{h_{s_{k + 1}}}{\mu g_{s_{k + 1}}} - 1\right) \leq
\sum^{k}_{i=1}h_{s_i}P^{pk}_{s_i} \leq W\left(\frac{h_{s_k}}{\mu
g_{s_k}} - 1\right), \ 1 \leq k \leq N - 1.
\end{equation}
If $q^{\ast}_{s_k} = \mu g_{s_k}P^{pk}_{s_k}$, $k = N$, then only
\eqref{pa_pa15} must be true and it follows that
\begin{equation}\label{pa_pa18}
\sum^{k}_{i=1}h_{s_i}P^{pk}_{s_i} \leq W\left(\frac{h_{s_k}}{\mu g_{s_k}}
- 1\right), \ k = N.
\end{equation}

\textbf{Lemma~6:} {\it There exists only one set of values for
$\{q^{\ast}_i\}$ that satisfies only one of the necessary
conditions \eqref{pa_pa12}, \eqref{pa_pa15} or \eqref{pa_pa16}.}

\textit{Proof:} \ It is equivalent to prove that there exists only
one $k$ that satisfies only one of \eqref{pa_pa14},
\eqref{pa_pa17} or \eqref{pa_pa18}. Let $L_j \triangleq
\sum^{j}_{i=1}h_{s_i} P^{pk}_{s_i}$ and $M_j \triangleq W(h_{s_j}
/ \mu g_{s_j} - 1)$ for brevity. Then it must be true that $L_0 <
L_1 < \cdots < L_N$, $M_1 > M_2 > \cdots > M_N$ and $L_0 < M_1$.
It can be seen that if \eqref{pa_pa18} holds, i.e., if $L_i <
M_i$, $\forall i$, $1 \leq i \leq N$, then \eqref{pa_pa14} and
\eqref{pa_pa17} do not hold.

If \eqref{pa_pa18} does not hold, then these exists such $l$ that
$L_i < M_i$, $\forall i$, $ 1 \leq i \leq l - 1$ and $L_i > M_i$,
$\forall i$, $1 \leq i \leq N$. The following two cases should be
considered. (i) If $L_{l - 1} < M_l < L_l$, \eqref{pa_pa14} holds
for $k = l$. Since $L_i < M_i$, $\forall i$, $1 \leq i \leq l -
1$, \eqref{pa_pa14} does not hold for $k < l$ as well. Since $M_i
< M_l < L_l \leq L_{i - 1}$, $\forall i$, $l + 1 \leq i$,
\eqref{pa_pa14} does not hold for $k > l$. Since $L_i < L_{i + 1}
< M_{i + 1}$, $\forall i$, $1 \leq i \leq l - 2$, \eqref{pa_pa17}
does not hold for $k < l - 1$. Since $L_{l - 1} < M_l$,
\eqref{pa_pa17} does not hold also for $k = l - 1$. Moreover,
since $M_i < L_i$, $\forall i$, $l \leq i$, \eqref{pa_pa17} does
not hold for $k > l - 1$. Therefore, only \eqref{pa_pa14} holds
for only $k = l$. (ii) If $M_l < L_{l - 1} < M_{l - 1}$,
\eqref{pa_pa17} holds for $k = l - 1$. Similar to the case (i), it
can be proved that only \eqref{pa_pa17} holds for only $k = l -
1$. 
$\hfill\square$

Using Lemma~6, Algorithm~2 is developed to find the unique $k$ in
Lemma~5.
\begin{algorithm}
\caption{Algorithm for finding $k$ in Lemma~5} \label{pa_pa_al}
\begin{algorithmic}
\STATE \textbf{Initialize:} $k = 0$, $c = 0$

\WHILE{$c = 0$}

\STATE $k = k + 1$

\IF {$\sum^{k - 1}_{i=1}h_{s_i}P^{pk}_{s_i} < W(h_{s_k} / \mu
g_{s_k} - 1) < \sum^{k}_{i=1}h_{s_i} P^{pk}_{s_i}$}

\STATE $c = 1$

\ENDIF

\IF {$\{W(h_{s_{k + 1}} / \mu g_{s_{k + 1}} - 1) \leq
\sum^{k}_{i=1} h_{s_i}P^{pk}_{s_i} \leq W(h_{s_k} / \mu g_{s_k} -
1)$ and $k \leq N - 1\}$ or
\\ $\{\sum^{k}_{i=1}h_{s_i} P^{pk}_{s_i} \leq W(h_{s_k} / \mu
g_{s_k} - 1)$ and $k = N\}$}

\STATE $c = 2$

\ENDIF

\ENDWHILE

\STATE \textbf{Output:} $k$, $c$
\end{algorithmic}
\end{algorithm}
Note that $k$ satisfies \eqref{pa_pa14} and \eqref{pa_pa17} or
\eqref{pa_pa18} if the output of Algorithm \ref{pa_pa_al} is $c =
1$ and $c = 2$, respectively. Since $p^{\ast}_{s_i} =
q^{\ast}_{s_i} / \mu g_{s_i}$, when $c = 1$, we obtain
\begin{equation}\label{pa_pa_sln1}
p^{\ast}_{s_i} = \left\{ {\begin{array}{ll}
   P^{pk}_{s_i}, & 1 \leq i \leq k-1 \\
   W(1 / \mu g_{s_k} - 1 / h_{s_k}) - \sum^{k-1}_{i=1} h_{s_i}
   P^{pk}_{s_i} / h_{s_k}, & i = k \\
   0, & k+1 \leq i \leq N \\
\end{array}} \right. , \ \ 1 \leq i \leq N
\end{equation}
and when $c = 2$, we obtain
\begin{equation}\label{pa_pa_sln2}
p^{\ast}_{s_i} = \left\{ {\begin{array}{ll}
   P^{pk}_{s_i}, & 1 \leq i \leq k \\
   0, & k+1 \leq i \leq N \\
\end{array}} \right. , \ \ 1 \leq i \leq N.
\end{equation}

\subsection{Average transmit power with peak interference
power constraints}\label{pa_ap}

Consider $\mathcal{F^{\prime}} = \{\textrm{the \ constraints
\eqref{ic_pk} and \eqref{pc_av}}\}$. Then the dual function of the
problem \eqref{sc_eq} can be written as
\begin{equation}\label{pa_ap_dual}
f_4(\{\lambda_i\}) \triangleq \textrm{E} \left\{ f^{\prime}_4
(\bm{h}, \bm{g}) \right\} + \sum^{N}_{i=1} \lambda_i P^{av}_i
\end{equation}
where $\{\lambda_i|1 \leq i \leq N\}$ are the nonnegative dual
variables associated with the corresponding constraints
\eqref{pc_av} and $f^{\prime}_4(\bm{h},\bm{g})$ is given by
\begin{subequations}
\begin{eqnarray}
f^{\prime}_4(\bm{h}, \bm{g}) \triangleq & & \max_{\{p_i(\bm{h},
\bm{g})\}} W \log \left( 1 + \frac{\sum^{N}_{i=1} h_i p_i (\bm{h},
\bm{g})} {W} \right) - \sum^{N}_{i = 1} \lambda_i p_i(\bm{h},
\bm{g}) \label{pa_ap1} \\
& & \textrm{s.t.} \ \sum^{N}_{i = 1} g_i p_i(\bm{h},\bm{g}) \leq
Q^{pk}. \label{pa_ap2}
\end{eqnarray}
\end{subequations}
Let $\{p^{\ast}_i\}$ denote the optimal solution of the problem
\eqref{pa_ap1}--\eqref{pa_ap2} where the dependence on $\bm{h}$
and $\bm{g}$ is dropped for brevity. The following three cases are
of interest.

\textit{Case 1:} Consider the case when $h_i \leq \lambda_i$,
$\forall i$. Similar to Case 1 in Section \ref{pa_pa}, it can be
seen from Lemma~3 that $p^{\ast}_i = 0$, $\forall i$.

\textit{Case 2:} Consider the case when $h_i \leq \lambda_i$ does
not hold for some $i$ and the constraint \eqref{pa_ap2} is
inactive at optimality. Let $(s_1, s_2, \cdots, s_N)$ denote a
permutation of the SU indexes such that $h_{s_1}/\lambda_{s_1} >
h_{s_2}/\lambda_{s_2} > \cdots > h_{s_N}/\lambda_{s_N}$. Since the
problem \eqref{pa_ap1}--\eqref{pa_ap2} without the constraint
\eqref{pa_ap2} has the same form as the problem \eqref{pa_aa1}, it
can be seen from \eqref{pa_aa_sln} that $p^{\ast}_{s_1} = W(1 /
\lambda_{s_1} - 1 / h_{s_1})$ and $p^{\ast}_{s_i} = 0$, $\forall
i$, $2 \leq i \leq N$, if it satisfies the constraint
\eqref{pa_ap2}, i.e., $\sum^{N}_{i = 1} g_{s_i}p^{\ast}_{s_i} =
g_{s_1}W(1 / \lambda_{s_1} - 1 / h_{s_1}) < Q^{pk}$.

\textit{Case 3:} Consider the case when $h_i \leq \lambda_i$ does
not hold for some $i$ and the constraint \eqref{pa_ap2} is active
at optimality, i.e., $g_{s_1}W(1 / \lambda_{s_1} - 1 / h_{s_1})
\geq Q^{pk}$. The dual function of the problem
\eqref{pa_ap1}--\eqref{pa_ap2} can be written as
$f_4^{\prime\prime}(\mu) \triangleq f^{\prime\prime\prime}_4 + \mu
Q^{pk}$, where $\mu$ is the nonnegative dual variable associated
with the constraint \eqref{pa_ap2}, and $f^{\prime\prime\prime}_4$
is given by
\begin{equation}\label{pa_ap3}
f^{\prime\prime\prime}_4 \triangleq \max_{\{p_i\}} W \log \left(1
+ \frac{\sum^{N}_{i=1} h_ip_i}{W}\right) - \sum^{N}_{i = 1}
\lambda_i p_i - \mu \sum^{N}_{i = 1} g_ip_i.
\end{equation}
Let $\mu^{\ast}$ denote the optimal dual variable. Also let
$F(\{p_i\})$ denote the objective function in the problem
\eqref{pa_ap3}. If $p^{\ast}_i
> 0$ for some $i$, the following must hold
\begin{equation}\label{pa_ap4}
\frac{\partial F(\{p_i\})}{\partial p_i}\bigg|_{\{p_i\} = \{
p^{\ast}_i \}} = \frac{h_i}{1 + \sum^{N}_{i=1} h_i p^{\ast}_i / W}
- \lambda_i - \mu^{\ast} g_i = 0.
\end{equation}
If $p^{\ast}_i = 0$ for some $i$, the following must hold
\begin{equation}\label{pa_ap5}
\frac{\partial F(\{p_i\})}{\partial p_i} \bigg|_{\{p_i\} =
\{p^{\ast}_i\}} = \frac{h_i}{1 + \sum^{N}_{i=1} h_ip^{\ast}_i / W}
- \lambda_i - \mu^{\ast}g_i \leq 0.
\end{equation}
Note that since the problem \eqref{pa_ap1}--\eqref{pa_ap2} is
convex, the necessary conditions \eqref{pa_ap4} and \eqref{pa_ap5}
for the optimal solution $\{p^{\ast}_i\}$ are also sufficient
conditions.

\textbf{Lemma~7:} {\it There exists at most two $j \neq k$ such
that $p^{\ast}_j > 0$ and $p^{\ast}_k > 0$.}

\textit{Proof:} \ We prove it by contradiction. It can be seen
from \eqref{pa_ap4} that if $p^{\ast}_i > 0$, $p^{\ast}_j > 0$,
and $p^{\ast}_k > 0$ for some $i \neq j$, $j \neq k$, $i \neq k$,
the following must hold
\begin{equation}\label{pa_ap6}
\frac{h_i}{\lambda_i + \mu^{\ast}g_i} = \frac{h_j}{\lambda_j +
\mu^{\ast}g_j} = \frac{h_k}{\lambda_k + \mu^{\ast}g_k}.
\end{equation}
Since $h_i$, $\lambda_i$, $g_i$, $h_j$, $\lambda_j$, $g_j$, $h_k$,
$\lambda_k$, and $g_k$ are independent constants given in the
problem \eqref{pa_ap1}--\eqref{pa_ap2}, and only $\mu^{\ast}$ is a
variable, \eqref{pa_ap6} can not be satisfied. $\hfill\square$

Lemma~7 shows that for the optimal power allocation under the
constraints \eqref{ic_pk} and \eqref{pc_av}, there exists at most
two users that transmit at nonzero power, while any other user
does not transmit.

Then Case 3 can be further divided into the following two subcases.

\textit{Case 3.1:} Consider the subcase when $p^{\ast}_k > 0$ and
$p^{\ast}_i = 0$, $\forall i \neq k$. Since the constraint
\eqref{pa_ap2} is active at optimality, i.e., $\sum^{N}_{i = 1}
g_ip^{\ast}_i = g_kp^{\ast}_k = Q^{pk}$, we obtain that
$p^{\ast}_k = Q^{pk} / g_k$. Then substituting $\{p^{\ast}_i\}$
into \eqref{pa_ap4} we have
\begin{equation}\label{pa_ap7}
\mu^{\ast} = \frac{1}{g_k / h_k + Q^{pk} / W} -
\frac{\lambda_k}{g_k}.
\end{equation}
Note that $\mu^{\ast}$ given in \eqref{pa_ap7} must satisfy
$\mu^{\ast} \geq 0$. Substituting $\{p^{\ast}_i\}$ into
\eqref{pa_ap5}, we can see that $\mu^{\ast}$ given in
\eqref{pa_ap7} also must satisfy
\begin{equation}\label{pa_ap8}
\mu^{\ast} \geq \frac{h_i / g_i}{1 + h_kQ^{pk} / g_kW} -
\frac{\lambda_i}{g_i}, \quad \forall \ i, i \neq k.
\end{equation}
Then Algorithm~3 can be used to find $k$.
\begin{algorithm}
\caption{Algorithm for finding $k$ in Case 3.1} \label{pa_ap_al1}
\begin{algorithmic}
\STATE $k = \textrm{arg}\max_{\{i\}} W \log\left(1 +
\frac{h_iQ^{pk}}{g_iW} \right) - \frac{\lambda_iQ^{pk}}{g_i}$
\STATE $\mu^{\ast} = \frac{1}{g_k / h_k + Q^{pk} / W} -
\frac{\lambda_k}{g_k}$ \IF {$\mu^{\ast} < \max_{\{i \neq k\}}
\frac{h_i / g_i}{1 + h_kQ^{pk} / g_kW} - \frac{\lambda_i}{g_i}$ or
$\mu^{\ast} < 0$} \STATE $k = 0$ \ENDIF \STATE \textbf{Output:}
$k$
\end{algorithmic}
\end{algorithm}
Note that $\{p_i^{\ast}\}$ does not exist in Case 3.1 if the
output of Algorithm \ref{pa_ap_al1} is $k = 0$.

\textit{Case 3.2:} Consider the subcase when $p^{\ast}_j > 0$,
$p^{\ast}_k > 0$, $j \neq k$ and $p^{\ast}_i = 0$, $\forall i, i
\neq j, i \neq k$. It follows from \eqref{pa_ap4} that
\begin{equation}\label{pa_ap9}
\frac{h_j}{\lambda_j + \mu^{\ast}g_j} = \frac{h_k}{\lambda_k +
\mu^{\ast}g_k}.
\end{equation}
Therefore, we obtain that
\begin{equation}\label{pa_ap10}
\mu^{\ast} = \frac{\lambda_j / h_j - \lambda_k / h_k}{g_k / h_k -
g_j / h_j}.
\end{equation}
Note that $\mu^{\ast}$ given in \eqref{pa_ap10} must satisfy
$\mu^{\ast} \geq 0$. Using \eqref{pa_ap4} and the fact that the
constraint \eqref{pa_ap2} is active at optimality, we have
\begin{equation}\label{pa_ap11}
\left\{ {\begin{array}{ll}
   h_jp^{\ast}_j + h_kp^{\ast}_k = W h_j / (\lambda_j + \mu^{\ast}
   g_j) - W\\
   g_jp^{\ast}_j + g_kp^{\ast}_k = Q^{pk}. \\
\end{array}} \right.
\end{equation}
Solving the system of equation \eqref{pa_ap11}, we obtain
\begin{equation}\label{pa_ap12}
p^{\ast}_j = \frac{Q^{pk} / g_k - a / h_k}{g_j / g_k - h_j / h_k},
\quad p^{\ast}_k = \frac{a / h_j - Q^{pk} / g_j}{h_k / h_j - g_k
/g_j}
\end{equation}
where $a \triangleq W h_j / (\lambda_j + \mu^{\ast}g_j) - W$. Note
that $p^{\ast}_j$ and $p^{\ast}_k$ given in \eqref{pa_ap12} must
satisfy $p^{\ast}_j > 0$ and $p^{\ast}_k > 0$. Substituting
$\{p^{\ast}_i\}$ and $\mu^{\ast}$ into \eqref{pa_ap5}, we can see
that $j$ and $k$ must satisfy
\begin{equation}\label{pa_ap13}
\frac{\lambda_j / h_j - \lambda_k / h_k}{g_k / h_k - g_j / h_j}
\geq \frac{\lambda_j / h_j - \lambda_i / h_i}{g_i / h_i - g_j /
h_j}, \quad \forall \ i, i \neq j, i \neq k.
\end{equation}
Then Algorithm~4 can be used to find $j$ and $k$.
\begin{algorithm}[h]
\caption{Algorithm for finding $j$ and $k$ in Case 3.2}
\label{pa_ap_al2}
\begin{algorithmic}
\STATE \textbf{Initialize:} $\mathcal{I} = \varnothing$

\FOR {$j = 1, \cdots, N-1$}

\FOR {$k = j + 1, \cdots, N$}

\STATE $\mu^{\ast} = \frac{\lambda_j / h_j - \lambda_k / h_k}{g_k
/ h_k - g_j / h_j}$

\IF {$\mu^{\ast} \geq 0$}

\STATE $a = W h_j / (\lambda_j + \mu^{\ast}g_j) - W$

\STATE $p^{\ast}_j = \frac{Q^{pk} / g_k - a / h_k}{g_j / g_k - h_j
/ h_k}$, $p^{\ast}_k = \frac{a / h_j - Q^{pk} / g_j}{h_k / h_j -
g_k / g_j}$

\IF {$p^{\ast}_j > 0$ and $p^{\ast}_k > 0$}

\STATE $\mathcal{I} = \mathcal{I} \cup \{(j,k)\}$

\STATE $v_{j,k} = W \log\left(1 + \frac{h_jp^{\ast}_j + h_k
p^{\ast}_k}{W}\right) - \lambda_jp^{\ast}_j - \lambda_kp^{\ast}_k$

\ENDIF

\ENDIF

\ENDFOR

\ENDFOR

\STATE $(j,k) = \textrm{arg}\max_{\{(i,l) \in \mathcal{I}\}} v_{i,l}$

\IF {$\frac{\lambda_j / h_j - \lambda_k / h_k}{g_k / h_k - g_j /
h_j} < \max_{\{i \neq j,k\}}\frac{\lambda_j / h_j - \lambda_i /
h_i}{g_i / h_i - g_j / h_j}$}

\STATE $j = 0$, $k = 0$

\ENDIF

\STATE \textbf{Output:} $j$, $k$
\end{algorithmic}
\end{algorithm}
Note that $\{p_i^{\ast}\}$ does not exist if the output of
Algorithm~\ref{pa_ap_al2} is $j = 0$ and $k = 0$.

\subsection{Combinations of more than two power constraints}
\label{pa_papa}

Consider $\mathcal{F^{\prime}} = \{\textrm{the \ constraints
\eqref{pc_pk}, \eqref{pc_av}, and \eqref{ic_av}}\}$ or
$\mathcal{F^{\prime}} = \{\textrm{the \ constraints \eqref{ic_pk},
\eqref{pc_av}, and \eqref{ic_av}}\}$. It can be shown that the
corresponding dual functions of the problem \eqref{sc_eq} under
these two combinations of the power constraints have the same form
as those in Subsections~\ref{pa_pa} and \ref{pa_ap}, respectively.
Therefore, optimal solutions can be found similarly therein and,
thus, are omitted here.

Consider $\mathcal{F^{\prime}} = \{\textrm{the \ constraints
\eqref{pc_pk}, \eqref{ic_pk}, and \eqref{pc_av}}\}$ or
$\mathcal{F^{\prime}} = \{\textrm{the \ constraints \eqref{pc_pk},
\eqref{ic_pk}, and \eqref{ic_av}}\}$ or $\mathcal{F^{\prime}} =
\{\textrm{the \ constraints \eqref{pc_pk}, \eqref{ic_pk},
\eqref{pc_av}, and \eqref{ic_av}}\}$. It can be shown that the
corresponding dual functions of the problem \eqref{sc_eq} under
the first two combinations of the power constraints have the same
form as that \ under \ the \ third \ combination. \ Therefore, we
only focus on the case $\mathcal{F^{\prime}} = \{\textrm{the \
constraints$ $\eqref{pc_pk}, \eqref{ic_pk}, \eqref{pc_av}, and
\eqref{ic_av}}\}$. Then the dual function of the problem
\eqref{sc_eq} can be written as
\begin{equation}\label{pa_papa_dual}
f_5(\{\lambda_i\}, \mu) \triangleq \textrm{E}\left\{f^{\prime}_5
(\bm{h},\bm{g})\right\} + \sum^{N}_{i=1} \lambda_iP^{av}_i + \mu
Q^{av}
\end{equation}
where $\{\lambda_i|1 \leq i \leq N\}$ and $\mu$ are the
nonnegative dual variables associated with the corresponding
constraints in \eqref{pc_av} and \eqref{ic_av} and
$f^{\prime}_5(\bm{h},\bm{g})$ is given by
\begin{subequations}
\begin{eqnarray}
f^{\prime}_5(\bm{h}, \bm{g}) \triangleq \!\!\!\!\!\!\!\!\! & &
\max_{\{p_i(\bm{h}, \bm{g})\}} W \log \left(1 \!+\!
\frac{\sum^{N}_{i=1} h_i p_i(\bm{h}, \bm{g})}{W} \right) \!-\!
\sum^{N}_{i = 1} \lambda_i p_i(\bm{h}, \bm{g}) \!-\! \mu
\sum^{N}_{i = 1} g_ip_i(\bm{h},\bm{g}) \label{pa_papa1} \\
& & \textrm{s.t.} \ \sum^{N}_{i = 1} g_i p_i(\bm{h}, \bm{g})
\leq Q^{pk} \label{pa_papa2} \\
& & \ p_i(\bm{h},\bm{g}) \leq P^{pk}_i, \ \forall \ i.
\label{pa_papa3}
\end{eqnarray}
\end{subequations}
Let $\{p^{\ast}_i\}$ denote the optimal solution of the problem
\eqref{pa_papa1}--\eqref{pa_papa3} where the dependence on
$\bm{h}$ and $\bm{g}$ is dropped for brevity. The following cases
are of interest.

\textit{Case 1:} Consider the case when $h_i \leq \lambda_i + \mu
g_i$, $\forall i$. Similar to Case 1 in Subsections~\ref{pa_pa}
and \ref{pa_ap}, it can be seen from Lemma~3 that $p^{\ast}_i =
0$, $\forall i$.

\textit{Case 2:} Consider the case when $h_i \leq \lambda_i + \mu
g_i$ does not hold for some $i$ and the constraint
\eqref{pa_papa2} is inactive at optimality. Since the problem
\eqref{pa_papa1}--\eqref{pa_papa3} without the constraint
\eqref{pa_papa2} has the same form as the problem
\eqref{pa_pa1}--\eqref{pa_pa2}, $\{p^{\ast}_i\}$ can be found
using Algorithm \ref{pa_pa_al} and \eqref{pa_pa_sln1} or
\eqref{pa_pa_sln2} if it satisfies the constraint
\eqref{pa_papa2}.

\textit{Case 3:} Consider the case when $h_i \leq \lambda_i + \mu
g_i$ does not hold for some $i$ and the constraint
\eqref{pa_papa2} is active at optimality. The dual function of the
problem \eqref{pa_papa1}--\eqref{pa_papa3} can be written as
$f^{\prime\prime}_5(\beta) \triangleq f^{\prime\prime\prime}_5 +
\beta Q^{pk}$, where $\beta$ is the nonnegative dual variable
associated with the constraint \eqref{pa_papa2} and
$f^{\prime\prime\prime}_5$ is given by
\begin{subequations}
\begin{eqnarray}
f^{\prime\prime\prime}_5 \triangleq & & \max_{\{p_i\}} W \log
\left( 1 + \frac{\sum^{N}_{i=1} h_i p_i}{W}\right) -
\sum^{N}_{i = 1} \gamma_i p_i - \beta \sum^{N}_{i = 1}g_i p_i
\label{pa_papa4} \\
& & \textrm{s.t.} \ p_i \leq P^{pk}_i, \ \forall \ i.
\label{pa_papa5}
\end{eqnarray}
\end{subequations}
where $\gamma_i \triangleq \lambda_i + \mu g_i$. Let
$\beta^{\ast}$ denote the optimal dual variable and $F(\{p_i\})$
stands for the objective function in the problem \eqref{pa_papa4}.
If $P^{pk}_i > p^{\ast}_i
> 0$ for some $i$, the following must hold
\begin{equation}\label{pa_papa6}
\frac{\partial F(\{p_i\})}{\partial p_i}\bigg|_{\{p_i\} =
\{p^{\ast}_i\}} = \frac{h_i}{1 + \sum^{N}_{i=1} h_i p^{\ast}_i /
W} - \gamma_i - \beta^{\ast}g_i = 0.
\end{equation}
If $p^{\ast}_i = P^{pk}_i$ for some $i$, the following must hold
\begin{equation}\label{pa_papa7}
\frac{\partial F(\{p_i\})}{\partial p_i}\bigg|_{\{p_i\} =
\{p^{\ast}_i\}} = \frac{h_i}{1 + \sum^{N}_{i=1} h_ip^{\ast}_i / W}
- \gamma_i - \beta^{\ast}g_i \geq 0.
\end{equation}
Moreover, if $p^{\ast}_i = 0$ for some $i$, the following must hold
\begin{equation}\label{pa_papa8}
\frac{\partial F(\{p_i\})}{\partial p_i}\bigg|_{\{p_i\} =
\{p^{\ast}_i\}} = \frac{h_i}{1 + \sum^{N}_{i=1} h_ip^{\ast}_i / W}
- \gamma_i - \beta^{\ast}g_i \leq 0.
\end{equation}
Note that since the problem \eqref{pa_papa1}--\eqref{pa_papa3} is
convex, the necessary conditions \eqref{pa_papa6},
\eqref{pa_papa7} and \eqref{pa_papa8} for the optimal solution
$\{p^{\ast}_i\}$ are also sufficient conditions.

\textbf{Lemma~8:} {\it There exists at most two $j$ and $k$, $j
\neq k$ such that $P^{pk}_j > p^{\ast}_j > 0$ and $P^{pk}_k >
p^{\ast}_k
> 0$.}

\textit{Proof:} \ We prove it by contradiction. It can be seen
from \eqref{pa_papa6} that if $P^{pk}_i > p^{\ast}_i > 0$,
$P^{pk}_j > p^{\ast}_j > 0$, and $P^{pk}_k > p^{\ast}_k > 0$ for
some $i \neq j$, $j \neq k$, $i \neq k$, the following must be
true
\begin{equation}\label{pa_papa9}
\frac{h_i}{\gamma_i + \beta^{\ast}g_i} = \frac{h_j}{\gamma_j +
\beta^{\ast}g_j} = \frac{h_k}{\gamma_k + \beta^{\ast}g_k}.
\end{equation}
Since $h_i$, $\gamma_i$, $g_i$, $h_j$, $\gamma_j$, $g_j$, $h_k$,
$\gamma_k$, and $g_k$ are independent constants given in the
problem \eqref{pa_papa1}--\eqref{pa_papa3}, and only
$\beta^{\ast}$ is a variable, \eqref{pa_papa9} can not be
satisfied. $\hfill\square$

Lemma~8 shows that for the optimal power allocation under the
constraints \eqref{pc_pk}, \eqref{ic_pk}, \eqref{pc_av} and
\eqref{ic_av}, there exists at most two user that transmit at
nonzero power and below their peak power, while any other user
either does not transmit or transmits at its peak power.

Then Case 3 can be further divided into the following two
subcases.

\textit{Case 3.1:} Consider the subcase when $P^{pk}_k >
p^{\ast}_k > 0$ and $p^{\ast}_i \in \{P^{pk}_i, 0\}$, $\forall i
\neq k$. Let $\mathcal{N}_1$ and $\mathcal{N}_0$ denote the sets
of SU indexes such that $p^{\ast}_i = P^{pk}_i$, $\forall i \in
\mathcal{N}_1$ and $p^{\ast}_i = 0$, $\forall i \in
\mathcal{N}_0$. Since the constraint \eqref{pa_papa2} is active at
optimality, i.e., $\sum^{N}_{i = 1} g_ip^{\ast}_i = g_kp^{\ast}_k
+ \sum_{i \in \mathcal{N}_1} g_iP^{pk}_i = Q^{pk}$, we obtain
$p^{\ast}_k = (Q^{pk} - \sum_{i \in \mathcal{N}_1} g_iP^{pk}_i) /
g_k$. Note that $p^{\ast}_k$ given here must satisfy $P^{pk}_k >
p^{\ast}_k > 0$. Then substituting $\{p^{\ast}_i\}$ into
\eqref{pa_papa6} we obtain
\begin{equation}\label{pa_papa10}
\beta^{\ast} = \frac{h_k / g_k}{1 + \left(h_k(Q^{pk} - \sum_{i \in
\mathcal{N}_1} g_iP^{pk}_i) / g_k + \sum_{i \in \mathcal{N}_1}
h_iP^{pk}_i\right) / W} - \frac{\gamma_k}{g_k}.
\end{equation}
Note that $\beta^{\ast}$ given by \eqref{pa_papa10} must satisfy
$\beta^{\ast} \geq 0$. Substituting $\{p^{\ast}_i\}$ into
\eqref{pa_papa7} we can see that $\beta^{\ast}$ given by
\eqref{pa_papa10} must satisfy
\begin{equation}\label{pa_papa11}
\beta^{\ast} \leq \frac{h_i / g_i}{1 + \left(h_k(Q^{pk} - \sum_{i
\in \mathcal{N}_1} g_iP^{pk}_i) / g_k + \sum_{i \in \mathcal{N}_1}
h_iP^{pk}_i\right) / W} - \frac{\gamma_i}{g_i}, \ \forall \ i \in
\mathcal{N}_1.
\end{equation}
Substituting $\{p^{\ast}_i\}$ into \eqref{pa_papa8}, we can see
that $\beta^{\ast}$ given in \eqref{pa_papa10} also must satisfy
\begin{equation}\label{pa_papa12}
\beta^{\ast} \geq \frac{h_i / g_i}{1 + \left(h_k (Q^{pk} - \sum_{i
\in \mathcal{N}_1} g_i P^{pk}_i) / g_k + \sum_{i \in
\mathcal{N}_1} h_i P^{pk}_i\right) / W} - \frac{\gamma_i}{g_i}, \
\forall \ i \in \mathcal{N}_0.
\end{equation}
Let $\mathcal{S}^{(1)}_{i}, \mathcal{S}^{(2)}_{i}, \cdots,
\mathcal{S}^{(2^{N - 1})}_{i}$ denote all the subsets of the set
$\mathcal{N} \backslash \{i\}$ where $\backslash$ denotes the set
difference operator. Then Algorithm~5 can be used to find $k$,
$\mathcal{N}_1$, and $\mathcal{N}_0$.
\begin{algorithm}
\caption{Algorithm for finding $k$, $\mathcal{N}_1$,
$\mathcal{N}_0$ in Case 3.1} \label{pa_papa_al1}
\begin{algorithmic}
\STATE \textbf{Initialize:} $\mathcal{I} = \varnothing$ \FOR {$k =
1, 2, \cdots, N$} \FOR {$l = 1, 2, \cdots, 2^{N - 1}$}

\STATE $\mathcal{N}_1 = \mathcal{S}^{(l)}_{k}$

\STATE $p^{\ast}_k = (Q^{pk} - \sum_{i \in \mathcal{N}_1} g_i
P^{pk}_i) / g_k$

\IF {$P^{pk}_k > p^{\ast}_k > 0$}

\STATE $\mathcal{I} = \mathcal{I} \cup \{l\}$

\STATE $r_l = W \log\left(1 + \frac{h_kp^{\ast}_k + \sum_{i \in
\mathcal{N}_1} h_iP^{pk}_i}{W}\right) - \gamma_kp^{\ast}_k -
\sum_{i \in \mathcal{N}_1} \gamma_iP^{pk}_i$

\ENDIF

\ENDFOR

\STATE $v_k = \max_{\{i \in \mathcal{I}\}} r_i$, $t = \textrm{arg}
\max_{\{i \in \mathcal{I}\}} r_i$

\STATE $\mathcal{S}^{\ast}_{k} = \mathcal{S}^{(t)}_{k}$

\STATE $\mathcal{I} = \varnothing$

\ENDFOR

\STATE $k = \textrm{arg} \max_{\{i\}}v_i$

\STATE $\mathcal{N}_1 = \mathcal{S}^{\ast}_{k}$

\STATE $\mathcal{N}_0 = \mathcal{N} \backslash \mathcal{N}_1
\backslash \{k\}$

\STATE $\beta^{\ast} = \frac{h_k / g_k}{1 + \left(h_k(Q^{pk} -
\sum_{i \in \mathcal{N}_1} g_iP^{pk}_i) / g_k + \sum_{i \in
\mathcal{N}_1} h_iP^{pk}_i\right) / W} - \frac{\gamma_k}{g_k}$

\IF {$\beta^{\ast} < 0$ or $\beta^{\ast} > \frac{h_i / g_i}{1 +
\left(h_k(Q^{pk} - \sum_{i \in \mathcal{N}_1} g_i P^{pk}_i) / g_k
+ \sum_{i \in \mathcal{N}_1} h_iP^{pk}_i\right) / W} -
\frac{\gamma_i}{g_i}, \exists i \in \mathcal{N}_1$ \\ or
$\beta^{\ast} < \frac{h_i / g_i}{1 + \left(h_k (Q^{pk} - \sum_{i
\in \mathcal{N}_1} g_i P^{pk}_i) / g_k + \sum_{i \in
\mathcal{N}_1} h_i P^{pk}_i\right) / W} - \frac{\gamma_i}{g_i},
\exists i \in \mathcal{N}_0$ }

\STATE $k = 0$

\ENDIF

\STATE \textbf{Output:} $k$, $\mathcal{N}_1$, $\mathcal{N}_0$
\end{algorithmic}
\end{algorithm}
Note that $\{p_i^{\ast}\}$ does not exist if the output of
Algorithm~\ref{pa_papa_al1} is $k = 0$.

\textit{Case 3.2:} Consider the subcase when $P^{pk}_j >
p^{\ast}_j > 0$, $P^{pk}_k > p^{\ast}_k > 0$ and $p^{\ast}_i \in
\{P^{pk}_i, 0\}$, $\forall i \neq j, k$. Let $\mathcal{N}_1$ and
$\mathcal{N}_0$ denote the sets of SU indexes such that
$p^{\ast}_i = P^{pk}_i$, $\forall i \in \mathcal{N}_1$ and
$p^{\ast}_i = 0$, $\forall i \in \mathcal{N}_0$, respectively. It
follows from \eqref{pa_papa6} that
\begin{equation}\label{pa_papa13}
\frac{h_j}{\gamma_j + \beta^{\ast}g_j} = \frac{h_k}{\gamma_k +
\beta^{\ast}g_k}.
\end{equation}
Therefore, we obtain that
\begin{equation}\label{pa_papa14}
\beta^{\ast} = \frac{\gamma_j / h_j - \gamma_k / h_k}{g_k / h_k -
g_j / h_j}.
\end{equation}
Note that $\beta^{\ast}$ given in \eqref{pa_papa14} must satisfy
$\beta^{\ast} \geq 0$. Following \eqref{pa_papa6} and the fact
that the constraint \eqref{pa_papa2} is active at optimality, we
have
\begin{equation}\label{pa_papa15}
\left\{ {\begin{array}{ll}
   h_jp^{\ast}_j + h_kp^{\ast}_k = W h_j / (\gamma_j +
   \beta^{\ast}g_j) - W - \sum_{i \in \mathcal{N}_1} h_iP^{pk}_i \\
   g_jp^{\ast}_j + g_kp^{\ast}_k = Q^{pk} - \sum_{i \in
   \mathcal{N}_1} g_iP^{pk}_i.\\
\end{array}} \right.
\end{equation}
Solving the system of equation \eqref{pa_papa15}, we obtain
\begin{equation}\label{pa_papa16}
p^{\ast}_j = \frac{a / g_k - b / h_k}{g_j / g_k - h_j / h_k},
\quad p^{\ast}_k = \frac{b / h_j - a / g_j}{h_k / h_j - g_k / g_j}
\end{equation}
where $a \triangleq Q^{pk} - \sum_{i \in \mathcal{N}_1}
g_iP^{pk}_i$ and $b \triangleq W h_j / (\gamma_j +
\beta^{\ast}g_j) - W - \sum_{i \in \mathcal{N}_1} h_iP^{pk}_i$.
Note that $p^{\ast}_j$ and $p^{\ast}_k$ given in \eqref{pa_papa16}
must satisfy $P^{pk}_j > p^{\ast}_j > 0$ and $P^{pk}_k >
p^{\ast}_k > 0$. Substituting $\{p^{\ast}_i\}$ and $\beta^{\ast}$
given by \eqref{pa_papa14} into \eqref{pa_papa7}, we obtain
\begin{equation}\label{pa_papa17}
\frac{\gamma_j / h_j - \gamma_k / h_k}{g_k / h_k - g_j / h_j} \leq
\frac{\gamma_j / h_j - \gamma_i / h_i}{g_i / h_i - g_j / h_j},
\quad \forall \ i \in \mathcal{N}_1.
\end{equation}
Moreover, substituting $\{p^{\ast}_i\}$ and $\beta^{\ast}$ given
by \eqref{pa_papa14} into \eqref{pa_papa8}, we also obtain
\begin{equation}\label{pa_papa18}
\frac{\gamma_j / h_j - \gamma_k / h_k}{g_k / h_k - g_j / h_j} \geq
\frac{\gamma_j / h_j - \gamma_i / h_i}{g_i / h_i - g_j / h_j},
\quad \forall \ i \in \mathcal{N}_0.
\end{equation}
Let $\mathcal{S}^{(1)}_{i,j}, \mathcal{S}^{(2)}_{i,j}, \cdots,
\mathcal{S}^{(2^{N - 2})}_{i,j}$ denote all the subsets of the set
$\mathcal{N} \backslash \{i,j\}$. Then Algorithm~6 can be used to
find $j$, $k$, $\mathcal{N}_1$, and $\mathcal{N}_0$.
\begin{algorithm}
\caption{Algorithm for finding $j$, $k$, $\mathcal{N}_1$,
$\mathcal{N}_0$ in Case 3.2} \label{pa_papa_al2}
\begin{algorithmic}
\STATE \textbf{Initialize:} $\mathcal{I} = \varnothing$

\FOR {$j = 1, 2, \cdots, N-1$}

\FOR {$k = j + 1, \cdots, N$}

\FOR {$l = 1, 2, \cdots, 2^{N - 2}$}

\STATE $\mathcal{N}_1 = \mathcal{S}^{(l)}_{j,k}$

\STATE $\beta^{\ast} = \frac{\gamma_j / h_j - \gamma_k / h_k}{g_k
/ h_k - g_j / h_j}$

\IF {$\beta^{\ast} \geq 0$}

\STATE $a \triangleq Q^{pk} - \sum_{i \in \mathcal{N}_1} g_i
P^{pk}_i$, $b \triangleq W h_j / (\gamma_j + \beta^{\ast}g_j) - W
- \sum_{i \in \mathcal{N}_1} h_iP^{pk}_i$

\STATE $p^{\ast}_j = \frac{a / g_k - b / h_k}{g_j / g_k - h_j /
h_k}$, $p^{\ast}_k = \frac{b / h_j - a / g_j}{h_k / h_j - g_k /
g_j}$

\IF {$P^{pk}_j > p^{\ast}_j > 0$ and $P^{pk}_k > p^{\ast}_k > 0$}

\STATE $\mathcal{I} = \mathcal{I} \cup \{l\}$

\STATE $r_l = W \log\left(1 + \frac{h_jp^{\ast}_j + h_kp^{\ast}_k
+ \sum_{i \in \mathcal{N}_1} h_iP^{pk}_i}{W} \right) -
\gamma_jp^{\ast}_j - \gamma_kp^{\ast}_k - \sum_{i \in
\mathcal{N}_1}\gamma_iP^{pk}_i$

\ENDIF

\ENDIF

\ENDFOR

\STATE $v_{j,k} = \max_{\{i \in \mathcal{I}\}} r_i$, $t =
\textrm{arg} \max_{\{i \in \mathcal{I}\}} r_i$

\STATE $\mathcal{S}^{\ast}_{j,k} = \mathcal{S}^{(t)}_{j,k}$

\STATE $\mathcal{I} = \varnothing$

\ENDFOR

\ENDFOR

\STATE $(j,k) = \textrm{arg} \max_{\{(i,l)\}}v_{i,l}$

\STATE $\mathcal{N}_1 = \mathcal{S}^{\ast}_{j,k}$

\STATE $\mathcal{N}_0 = \mathcal{N} \backslash \mathcal{N}_1
\backslash \{j,k\}$

\IF {$\frac{\gamma_j / h_j - \gamma_k / h_k}{g_k / h_k - g_j /
h_j} > \frac{\gamma_j / h_j - \gamma_i / h_i}{g_i / h_i - g_j /
h_j}, \exists i \in \mathcal{N}_1$ or $\frac{\gamma_j / h_j -
\gamma_k / h_k}{g_k / h_k - g_j / h_j} < \frac{\gamma_j / h_j -
\gamma_i / h_i}{g_i / h_i - g_j / h_j}, \exists i \in
\mathcal{N}_0$ }

\STATE $j = 0$, $k = 0$

\ENDIF

\STATE \textbf{Output:} $j$, $k$, $\mathcal{N}_1$, $\mathcal{N}_0$
\end{algorithmic}
\end{algorithm}
Note that $\{p_i^{\ast}\}$ does not exist if the output of
Algorithm~\ref{pa_papa_al2} is $j = 0$ and $k = 0$.

\section{Simulation Results}\label{sec_sim}

Consider a cognitive radio network which consists of one PU and
four SUs. For simplicity, we assume that only Rayleigh fading is
present in all links. The variance of the channel power gain is
set to $\sigma^2 = 1$. We also set $W = 1$, $P^{pk}_i = 10$,
$\forall i$, $P^{av}_i = 10$, $\forall i$, $Q^{pk} = 1$, and
$Q^{av} = 1$ as default values if no other values are specified
otherwise. The AWGN with unit PSD is assumed. We use 1000 randomly
generated channel power gains for $\bm{h}$ and $\bm{g}$ in our
simulations. The results are compared under the following five
combinations of the power constraints: the PTP with PIP
constraints (PTP+PIP), the PTP with AIP constraints (PTP+AIP), the
ATP with PIP constraints (ATP+PIP), the ATP with AIP constraints
(ATP+AIP), the PTP and ATP with PIP and AIP constraints
(PTP+ATP+PIP+AIP).

\begin{figure}[ht]
\begin{minipage}[b]{1.0\linewidth}
  \centering
  \centerline{\epsfig{figure=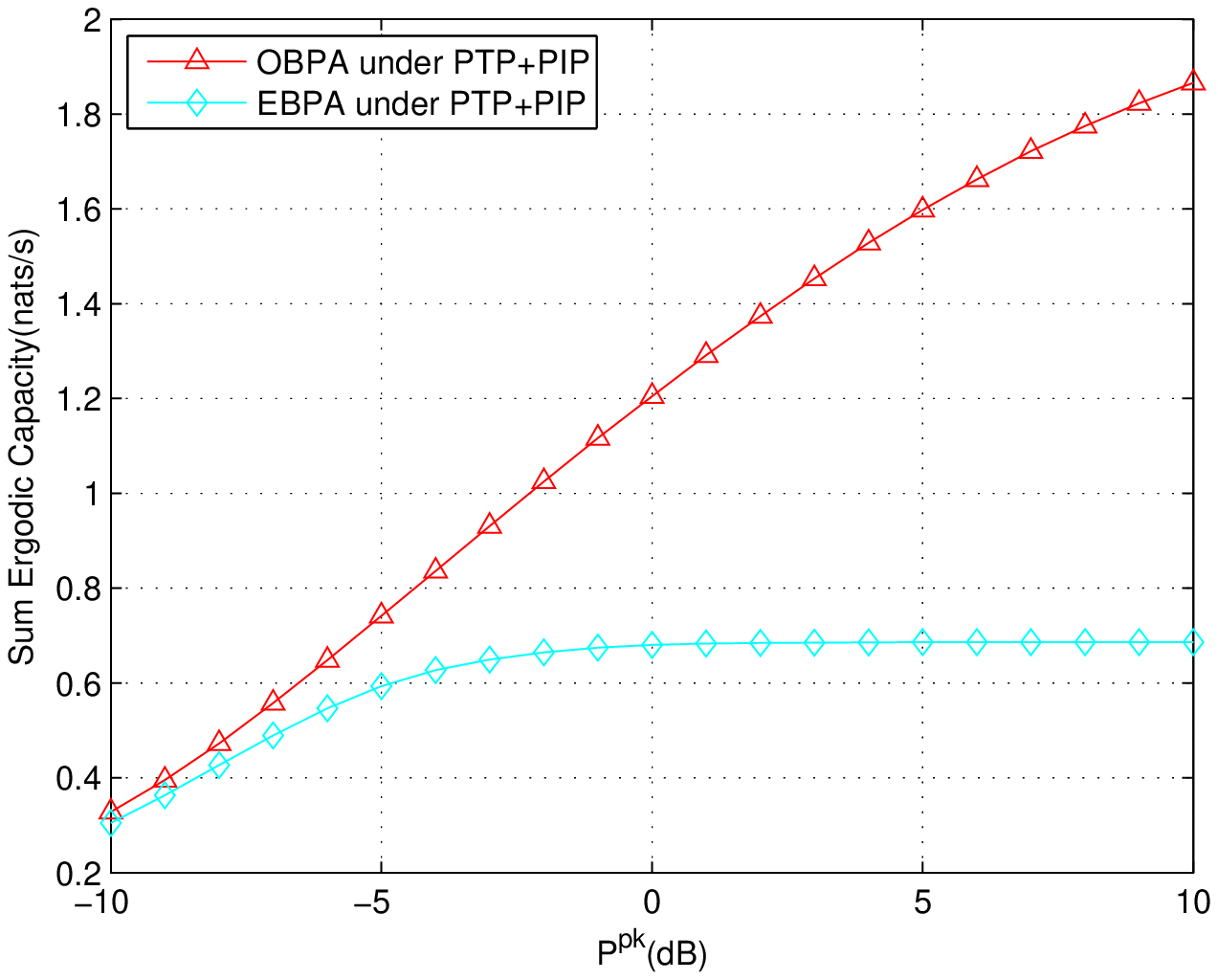,width=10.5cm}}
\end{minipage}
\caption{Sum ergodic capacity vs $P^{pk}$.} \label{fg_eqbp_p_pk}
\end{figure}

\begin{figure}[h]
\begin{minipage}[b]{1.0\linewidth}
  \centering
  \centerline{\epsfig{figure=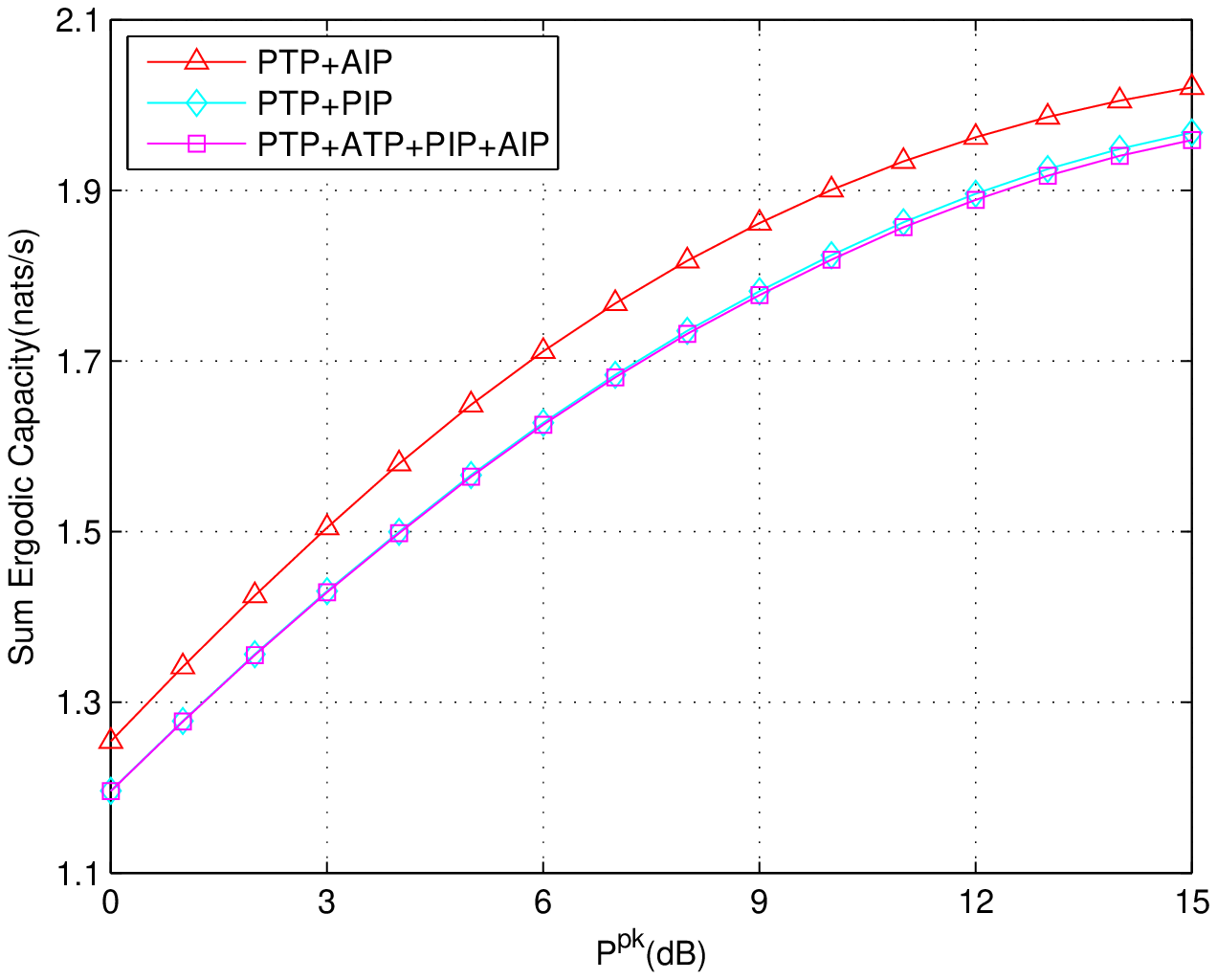,width=10.5cm}}
\end{minipage}
\caption{Sum ergodic capacity vs $P^{pk}$.} \label{fg_p_pk}
\end{figure}

First, we aim at showing by Fig.~\ref{fg_eqbp_p_pk} that the
information-theoretic limit for the sum ergodic capacity is indeed
significantly higher when bandwidth is allocated optimally as
compared to the case when it is allocated equally among SUs. In
this figure, OBPA stands for optimal bandwidth and power
allocation, while EBPA stands for equal bandwidth and power
allocation. The case of PTP+PIP is only depicted in
Fig.~\ref{fg_eqbp_p_pk}, but the conclusion about the superiority
of optimal bandwidth and power allocation holds true for other
combinations of power constraints. Then Fig.~\ref{fg_p_pk} shows
and compares the maximum sum ergodic capacity under PTP+PIP,
PTP+AIP and PTP+ATP+PIP+AIP constraints versus $P^{pk}$ where
$P^{pk} = P^{pk}_i$, $\forall i$ is assumed. It can be seen from
the figure that the maximum sum ergodic capacity achieved under
PTP+AIP is larger than that achieved under PTP+PIP for any given
$P^{pk}$. This is due to the fact that the AIP constraint is more
favorable than the PIP constraint from SUs' perspective, since the
former allows for more flexibility for SUs to allocate transmit
power over different channel fading states. It is also observed
that the performance under PTP+ATP+PIP+AIP is very close to that
under PTP+PIP that is because the PTP constraint dominates over
the ATP, PIP, and AIP constraints for all values of $P^{pk}$.

\begin{figure}[h]
\begin{minipage}[b]{1.0\linewidth}
  \centering
  \centerline{\epsfig{figure=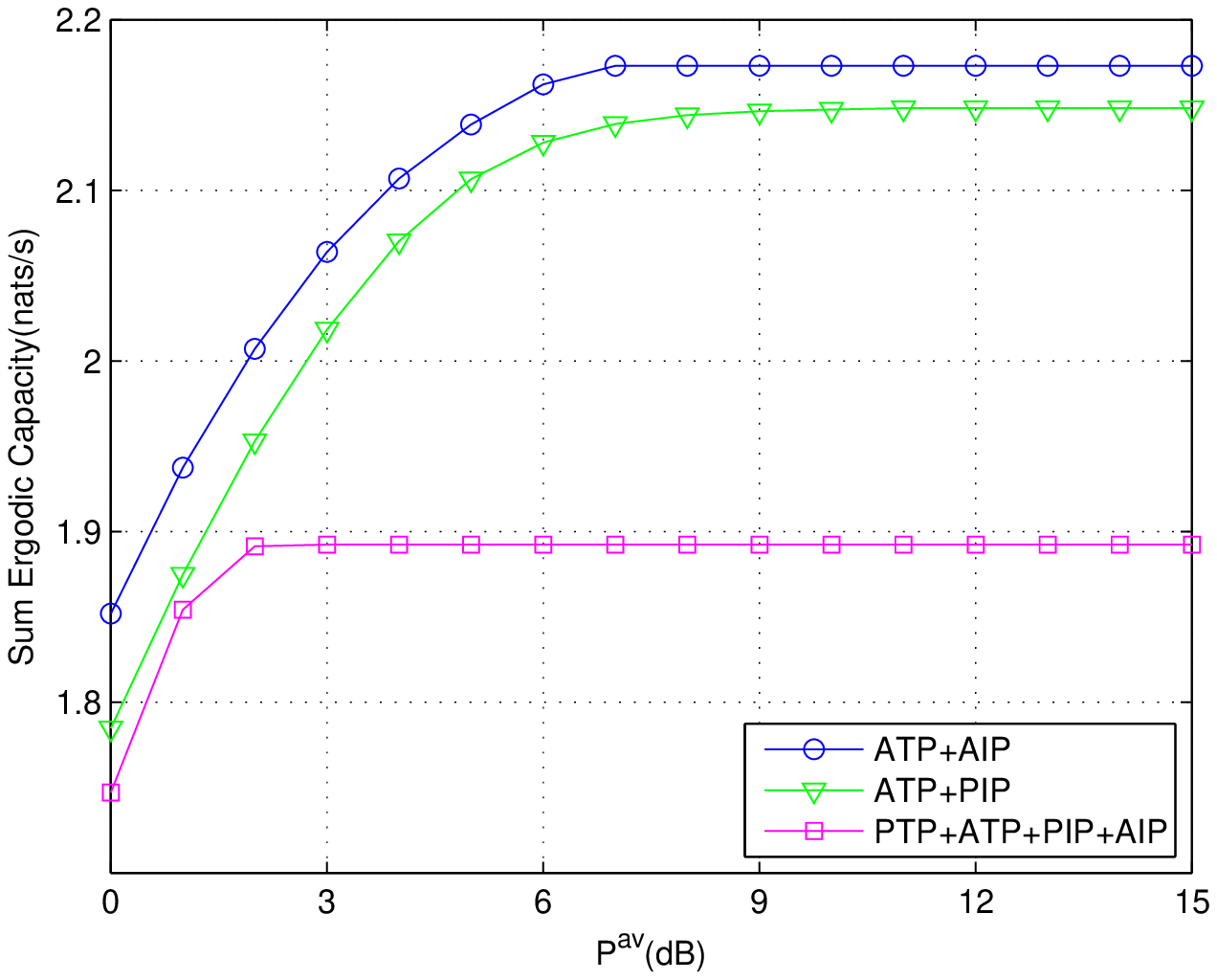,width=10.5cm}}
\end{minipage}
\caption{Sum ergodic capacity vs $P^{av}$.}\label{fg_p_av}
\end{figure}

Fig.~\ref{fg_p_av} shows the maximum sum ergodic capacity under
ATP+PIP, ATP+AIP and PTP+ATP+\\PIP+AIP constraints versus $P^{av}$
where $P^{av} = P^{av}_i$, $\forall i$ is assumed. The maximum
achievable sum ergodic capacity achieved under ATP+AIP is larger
than that achieved under ATP+PIP for all values of $P^{av}$ since
the PIP constraint is stricter than the AIP constraint. The sum
ergodic capacity under PTP+ATP+PIP+AIP is much smaller than that
under ATP+PIP and ATP+AIP due to the fact that the PTP constraint
is dominant over other constraints for all values of $P^{av}$.

\begin{figure}[h]
\begin{minipage}[b]{1.0\linewidth}
  \centering
  \centerline{\epsfig{figure=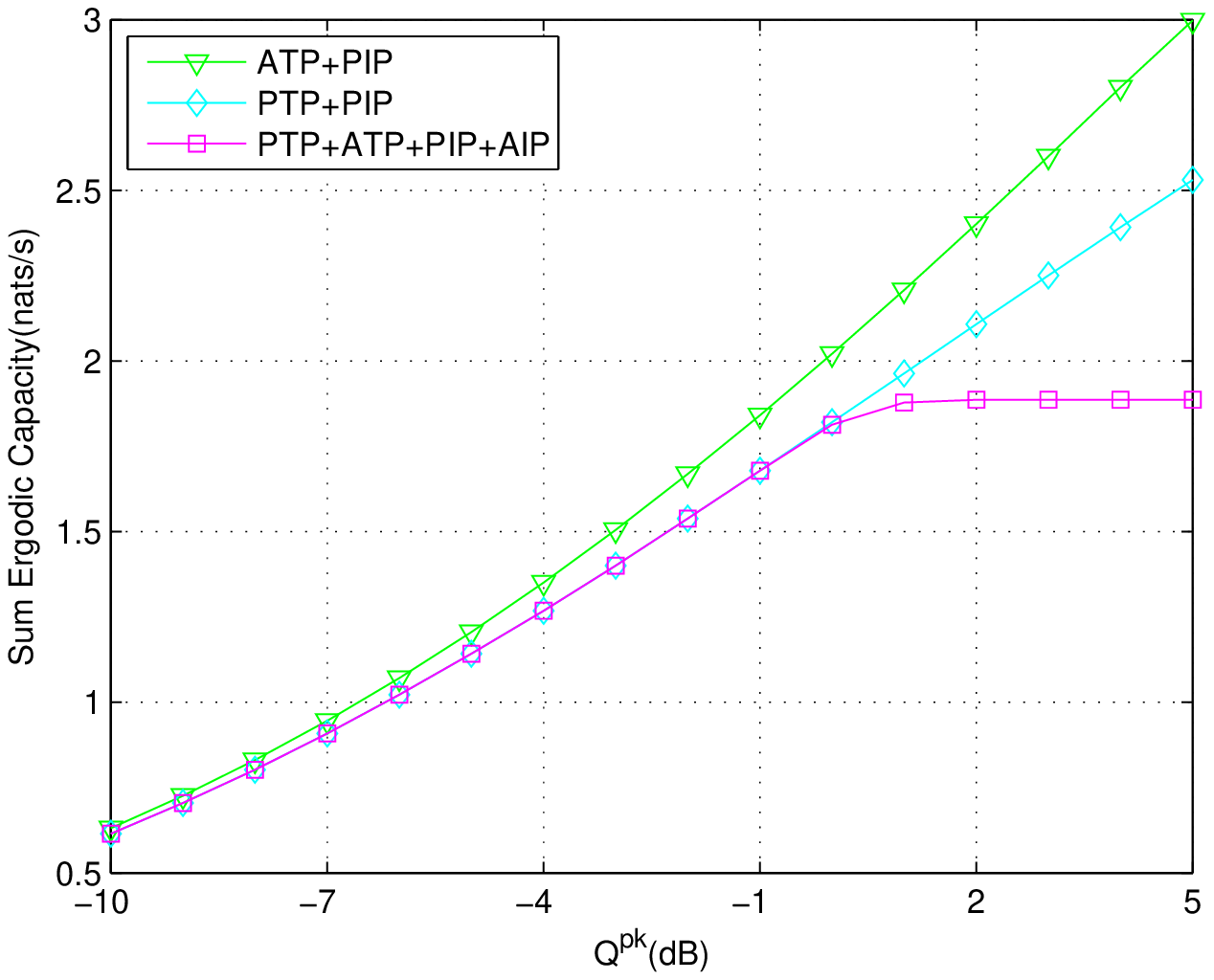,width=10.5cm}}
\end{minipage}
\caption{Sum ergodic capacity vs $Q^{pk}$.}\label{fg_q_pk}
\end{figure}

Fig.~\ref{fg_q_pk} shows the maximum sum ergodic capacity under
PTP+PIP, ATP+PIP and PTP+ATP+\\PIP+AIP constraints versus
$Q^{pk}$. It can be seen from the figure that the maximum sum
ergodic capacity achieved under ATP+PIP is larger than that
achieved under PTP+PIP for any given $Q^{pk}$. This is because the
power allocation is more flexible for SUs under the ATP constraint
than under the PTP constraint. The sum ergodic capacity under
PTP+ATP+PIP+AIP saturates earlier than that under PTP+PIP and
ATP+PIP, because it is restricted by the AIP constraint.

\begin{figure}[h]
\begin{minipage}[b]{1.0\linewidth}
  \centering
  \centerline{\epsfig{figure=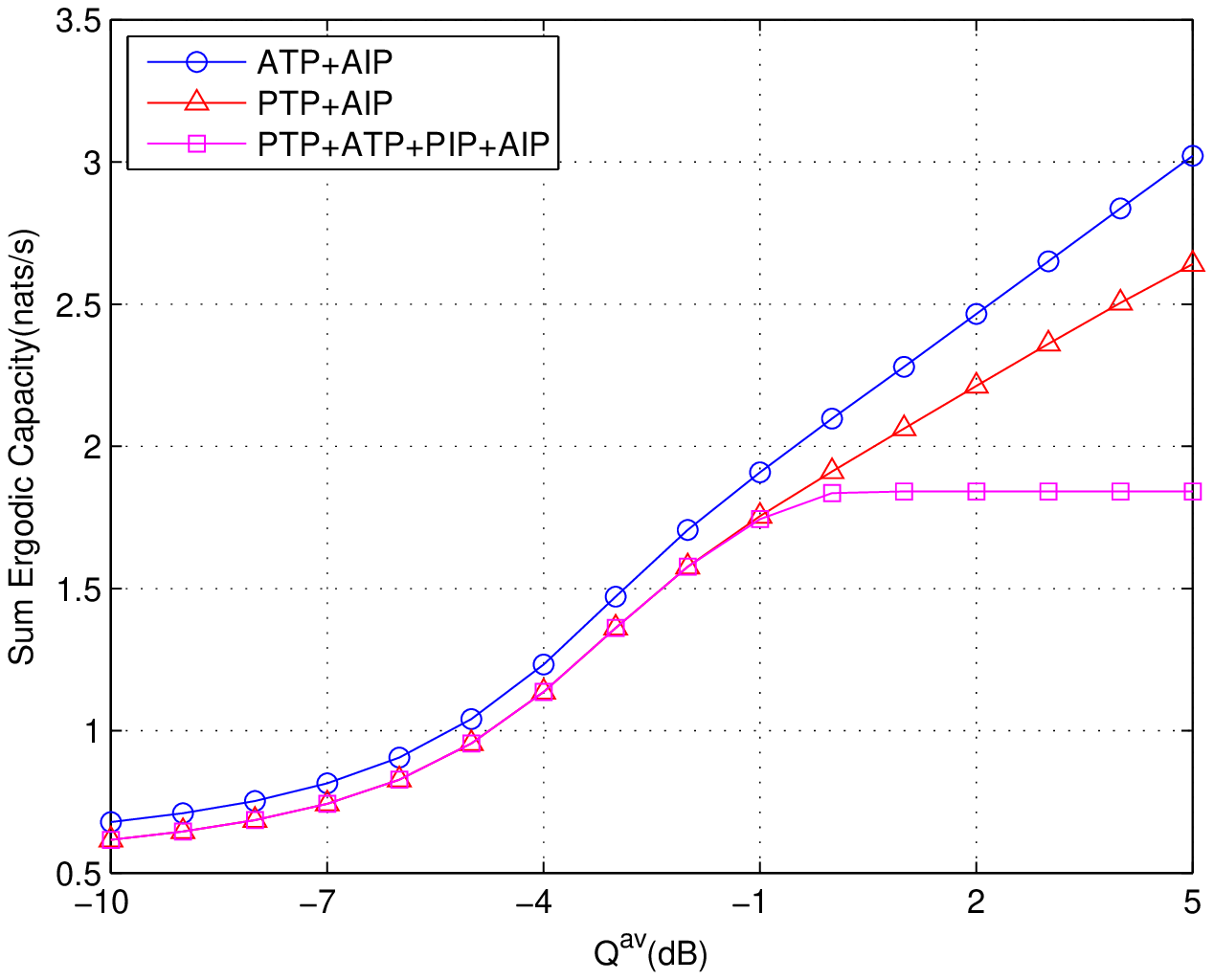,width=10.5cm}}
\end{minipage}
\caption{Sum ergodic capacity vs $Q^{av}$.}\label{fg_q_av}
\end{figure}

Fig.~\ref{fg_q_av} shows the maximum sum ergodic capacity under
PTP+AIP, ATP+AIP and PTP+ATP+\\PIP+AIP constraints versus
$Q^{av}$. Due to the same reasons as for the results in
Fig.~\ref{fg_q_pk}, the sum ergodic capacity achieved under
ATP+AIP is larger than that achieved under PTP+AIP. The sum
ergodic capacity under PTP+ATP+PIP+AIP saturates earlier than that
for PTP+AIP and ATP+AIP because of the presence of the PIP
constraint.

\begin{figure}[h]
\begin{minipage}[b]{1.0\linewidth}
  \centering
  \centerline{\epsfig{figure=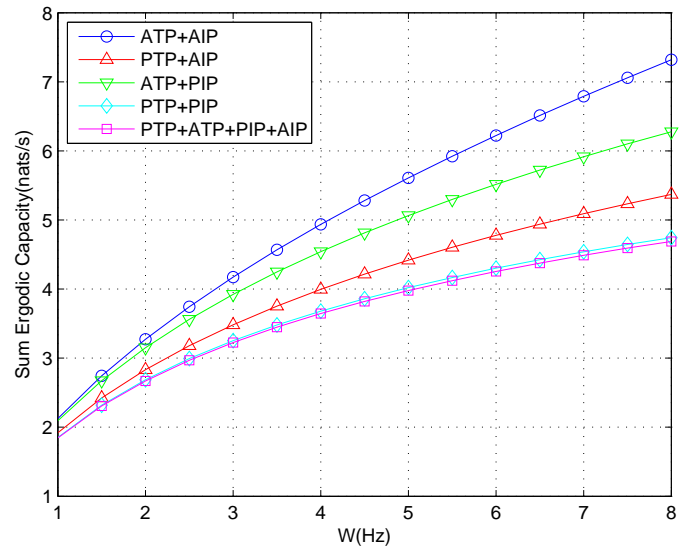,width=10.5cm}}
\end{minipage}
\caption{Sum ergodic capacity vs $W$.}\label{fg_w}
\end{figure}

Finally, Fig.~\ref{fg_w} shows the maximum sum ergodic capacity
under PTP+PIP, PTP+AIP, ATP+PIP, ATP+AIP and PTP+ATP+PIP+AIP
versus $W$. Similar performance comparison results as in the
previous figures can be observed. One difference is that the sum
ergodic capacities do not saturate with the increase of $W$.

\section{Conclusion}\label{sec_con}

A cognitive radio network where multiple SUs share the licensed
spectrum of a PU using the FDMA scheme has been considered. The
maximum achievable sum ergodic capacity of all the SUs has been
studied subject to the total bandwidth constraint of the licensed
spectrum and all possible combinations of the peak/average
transmit power constraints at the SUs and interference power
constraint imposed by the PU. Optimal bandwidth allocation has
been derived in each channel fading state for any given power
allocation. Using the structures of the optimal power allocations
under each combination of the power constraints, algorithms for
finding the optimal power allocations have been developed too.

\end{document}